\documentclass[twocolumn,secnumarabic,amssymb, nobibnotes, nofootinbib, prx,superscriptaddress]{revtex4-1}%-2
\usepackage[T1]{fontenc}
\usepackage[latin9]{inputenc}
\usepackage{dcolumn}% Align table columns on decimal point
\usepackage{siunitx}% Align table columns on decimal point
\usepackage{amsmath}
\usepackage{amssymb}
\usepackage{amsfonts}
\usepackage{graphicx}
\usepackage{hyperref}
\usepackage{color,soul}
\usepackage{mathrsfs}
\usepackage{multirow}
\usepackage{float}
\usepackage{bm}
\usepackage{dsfont}
\usepackage{txfonts}
\usepackage{multirow}
\usepackage[english]{babel}
\usepackage{tabularx}
\usepackage{xcolor}

\makeatletter
\DeclareMathOperator{\Tr}{Tr}
\newcolumntype{d}[1]{D{.}{.}{#1}} % for aligning table columns by decimal point
\newcommand\Tstrut{\rule{0pt}{2.4ex}}       % top strut
\newcommand\Bstrut{\rule[-1.3ex]{0pt}{0pt}} % bottom strut

\newcommand{\bra}[1]{\langle{#1}|}
\newcommand{\ket}[1]{|{#1}\rangle}

\newcommand{\beq}{\begin{equation}}
\newcommand{\eeq}{\end{equation}}
\newcommand{\bqa}{\begin{eqnarray}}
\newcommand{\eqa}{\end{eqnarray}}

\newcommand{\erf}[1]{Eq.~(\ref{#1})}

\newcommand{\eg}{{\em e.g.}}

\begin{document}
	
\title{Demonstration that Einstein-Podolsky-Rosen steering requires more than one bit of faster-than-light information transmission}

\author{Yu~Xiang}
\thanks{Y.X. and M.D.M. contributed equally to this work.}
\address{State Key Laboratory for Mesoscopic Physics, School of Physics, Frontiers Science Center for Nano-optoelectronics, $\&$ Collaborative Innovation Center of Quantum Matter, Peking University, Beijing 100871, China}
\address{Collaborative Innovation Center of Extreme Optics, Shanxi University, Taiyuan, Shanxi 030006, China}

\author{Michael D. Mazurek \textsuperscript{$\ast$}}
\email{michael.mazurek@nist.gov}
\address{Department of Physics, 390 UCB, University of Colorado, Boulder, Colorado 80309}
\address{National Institute of Standards and Technology, 325 Broadway, Boulder, Colorado 80305, USA}

\author{Joshua C. Bienfang}
\address{Joint Quantum Institute: University of Maryland and National Institute of Standards and Technology, 100 Bureau Drive, Gaithersburg MD 20899, USA}

\author{Michael A. Wayne}
\address{National Institute of Standards and Technology, 100 Bureau Drive, Gaithersburg MD 20899, USA}

\author{Carlos Abell\'an}
\address{ICFO - Institut de Ci\`encies Fot\`oniques, The Barcelona Institute of Science and Technology, 08860 Castelldefels (Barcelona), Spain}

\author{Waldimar Amaya}
\address{ICFO - Institut de Ci\`encies Fot\`oniques, The Barcelona Institute of Science and Technology, 08860 Castelldefels (Barcelona), Spain}

\author{Morgan W. Mitchell}
\address{ICFO - Institut de Ci\`encies Fot\`oniques, The Barcelona Institute of Science and Technology, 08860 Castelldefels (Barcelona), Spain}
\address{ICREA-Instituci\'o Catalana de Recerca i Estudis Avan\c cats, 08015 Barcelona, Spain}

\author{Richard P. Mirin}
\address{National Institute of Standards and Technology, 325 Broadway, Boulder, Colorado 80305, USA}

\author{Sae Woo Nam}
\address{National Institute of Standards and Technology, 325 Broadway, Boulder, Colorado 80305, USA}

\author{Qiongyi~He}
\email{qiongyihe@pku.edu.cn}
\address{State Key Laboratory for Mesoscopic Physics, School of Physics, Frontiers Science Center for Nano-optoelectronics, $\&$ Collaborative Innovation Center of Quantum Matter, Peking University, Beijing 100871, China}
\address{Collaborative Innovation Center of Extreme Optics, Shanxi University, Taiyuan, Shanxi 030006, China}

\author{Martin J. Stevens}
\address{National Institute of Standards and Technology, 325 Broadway, Boulder, Colorado 80305, USA}

\author{Lynden K. Shalm}
\address{National Institute of Standards and Technology, 325 Broadway, Boulder, Colorado 80305, USA}
\address{Department of Physics, 390 UCB, University of Colorado, Boulder, Colorado 80309}

\author{Howard M. Wiseman}
\email{h.wiseman@griffith.edu.au}
\address{Centre for Quantum Computation and Communication Technology (Australian Research Council), Centre for Quantum Dynamics, Griffith University, Brisbane QLD 4111, Australia}

\begin{abstract}

Schr\"odinger held that a local quantum system has some objectively real quantum state and no other (hidden) properties. He therefore took the Einstein-Podolsky-Rosen (EPR) phenomenon, which he generalized and called `steering', to require nonlocal wavefunction collapse.  Because this would entail faster-than-light (FTL) information transmission,  he doubted that it would be seen experimentally.  Here we report a demonstration of EPR steering with entangled photon pairs that puts---in Schr\"odinger's interpretation---a non-zero lower bound on the amount of FTL information transmission. We develop a family of $n$-setting loss-tolerant EPR-steering inequalities allowing for a size-$d$ classical message sent from Alice's laboratory to Bob's. For the case $n=3$ and $d=2$ (one bit) we observe a statistically significant violation. Our experiment closes the efficiency and locality loopholes, and we address the freedom-of-choice loophole by using quantum random number generators to independently choose Alice's and Bob's measurement basis settings. To close the efficiency and locality loopholes simultaneously, we introduce methods for quickly switching between three mutually unbiased measurement bases and for accurately characterizing the efficiency of detectors. From the space-time arrangement of our experiment, we can conclude that if the mechanism for the observed bipartite correlations is that Alice's measurement induces wave-function collapse of Bob's particle, then more than one bit of information must travel from Alice to Bob at more than three times the speed of light.

\end{abstract}
\maketitle

\section{Introduction}

In 1935, Einstein, Podolsky and Rosen (EPR) put forward their famous paradox to declare the quantum mechanical description of physical reality incomplete~\cite{EPR35}. Their argument relied on a remarkable feature of bipartite quantum entanglement, that the choice of measurement by one party (Alice) seems to affect, instantaneously, the type of state held by the second party (Bob). Einstein later called this phenomenon 
 ``spooky action-at-a-distance"~\cite{Einstein}. Meanwhile, in the same year as EPR,  Schr\"{o}dinger called the phenomenon ``steering" or ``piloting'' the remote state~\cite{Schrodinger35}, and discussed the possibility of using arbitrarily many types of measurement. Reid~\cite{Reid89} gave the first formulation of an experimental criterion for testing this effect based on EPR's example of position and momentum. More recently, Schr\"{o}dinger's generalisation was formalised,  as EPR steering, by one of us and co-workers~\cite{Howard07PRL,Howard07PRA}, as the experimental violation of the asymmetric model comprising a local hidden state (LHS) quantum model for Bob's system and a local hidden variable (LHV) model to generate Alice's measurement results. Subsequently, broad classes of experimental criteria to demonstrate EPR steering, known as steering inequalities, were derived~\cite{Eric09,Schneeloch13,Gerardo15,Otfried19}.
 Moreover, EPR-steering was shown~\cite{Howard07PRL,Howard07PRA} to be equivalent to the quantum information task of verifying entanglement in the absence of trust of Alice or her equipment. This is relevant for communication networks where reliability of devices and dishonest observers becomes an issue \cite{genuineCavalcanti,ANUexp,HowardOptica,CV-QKDexp}. 

While the largest impacts of EPR steering have been in its myriad quantum information applications ---including one-sided device-independent quantum cryptography~\cite{1sDIQKD,1sDIQKD_howard,ANUexp,HowardOptica,CV-QKDexp,YuQSS,GiannisQSS}, secure quantum teleportation with high fidelity~\cite{SQT13Reid,SQT15,SQT16_LiCM} and subchannel discrimination~\cite{subchannel1,subchannel2,subchannel3}---there has also been considerable theoretical and experimental interest in its foundational implications~\cite{Smith12,Bennet12,Wittmann12,Saunders12,Eric15,Fuwa15,Tischler18,DJS10,qjumps,Reid13,OneWayNatPhot,OneWayPryde,prlSu,hybrid18,science1,science2,He15,OGRMP}, as Einstein and Schr\"{o}dinger were concerned with. 
In particular, from the perspective of Schr\"{o}dinger, who was convinced of the completeness and correctness of the quantum state as a description of a {\em local} system, EPR steering implied a genuinely superluminal effect (which is why he doubted that it would be seen experimentally~\cite{Schrodinger35}). This effect was first demonstrated in a way that closed the locality and efficiency loopholes for EPR steering in 2012~\cite{Wittmann12}. However, that  experiment did not 
put any lower bound on {\em how much}  classical information would have to be sent faster-than-light (FTL) for Alice to steer Bob's state, in the absence of entanglement. 

The question of how to establish a lower bound on the apparent FTL information transmission in EPR steering was theoretically investigated in Refs.~\cite{sr_Nagy,pra_Sainz}. The authors showed that an infinite amount of classical communication (from an untrusted Alice to a trusted Bob), is necessary to simulate all EPR-steering correlations for any pure entangled state of two qubits. For any experimentally accessible correlations, the amount of classical communication required to violate the LHS model will be finite, and moreover is one measure of the strength of EPR steering. 

In this paper, we provide the first demonstration of EPR-steering correlations that provably require  {more than one bit} of FTL communication to simulate classically.  We do this by introducing and subsequently violating an inequality that is derived assuming one bit of FTL communication is possible. Our experiment addresses the efficiency, locality, and freedom-of-choice loopholes.
The requirements for experimentally testing our inequality while closing the efficiency and locality loopholes are many, but we highlight two which have interest outside this particular task.  
First, we achieve fast-switching between three mutually unbiased measurement bases with minimal losses by using a single Pockels cell. Second, we introduce a modification of Klyshko's method~\cite{Klyshko} 
to accurately characterize the efficiencies for Bob's two detectors.

This paper is structured as follows. In Sec.~\ref{sc:loopholes} we discuss how to close the loopholes in quantum nonlocality tests in general, in EPR-steering tests more particularly, and in our experiment allowing for one bit of FTL communication most particularly. In Sec.~\ref{sc:newInequality} we introduce a general family of inequalities for testing communication-assisted EPR steering, and in Sec.~\ref{sec:Ineqoct} we derive a specific inequality for the case when Bob can implement a set of three measurement settings.  Our derivation goes beyond the theory in Refs.~\cite{sr_Nagy,pra_Sainz} in that it includes the possibility of null results by Alice (as required for closing the efficiency loophole) and is optimized for the experimental conditions (Alice's efficiency, and the three measurements Bob can switch between). In Sec.~\ref{sc:experiment} we present an experimental test of our EPR-steering inequality using polarization-entangled photon pairs. Our experiment closes the locality, freedom-of-choice, and efficiency loopholes. We observe a statistically significant violation of our inequality, thus demonstrating correlations that---under the standard EPR-steering assumptions---require more than one bit of FTL information transmission.

\section{Closing loopholes in quantum nonlocality tests}\label{sc:loopholes}

There are four standard loopholes for tests of quantum nonlocality in a general sense: the locality loophole, the freedom-of-choice loophole, the efficiency (or detection) loophole, and the memory loophole~ \cite{larsson14}. These loopholes are best known for experimental tests of Bell-nonlocality with two parties, where they have the following meaning.  The \textit{locality loophole} is closed if and only if each party's choice of setting is space-like separated from the other party's observation of the measurement result. The \textit{freedom-of-choice loophole} refers to the requirement that the setting choices by the two parties are `free', i.e. not influenced by any hidden variables \cite{Bell, AbellanNature2018}.
In principle, the closure of this loophole can never be absolutely assured, but in practice each party can use a state-of-the-art random number generator for this purpose {~\cite{abellan15}}. 
Closing the \textit{efficiency loophole} requires that all runs of the experiment are included in the data analysis, not just a subensemble where both particles are detected, for example~\cite{Pearle}. This can be achieved by having sufficiently high detection efficiency on both sides.  Finally, the \textit{memory loophole} applies to the analysis of finite experimental data. If all trials are assumed to be independent and identical it is possible for an adversary to use information (memory) about past trials to try and influence or bias the outcomes of future trials. This can lead to the assignment of greater confidence to a violation than is warranted~\cite{larsson14}. All four loopholes have been simultaneously closed in a number of LHF Bell experiments~\cite{LHFnature,LHFprl1,LHFprl2,rosenfeld17,li18}.

EPR steering has the same loopholes, but asymmetrically. Because Bob's apparatus is trusted, {\em efficiency} only matters on Alice's side; it is permissible to consider the sub-ensemble where Bob detects a particle. Moreover, under some conditions it is only necessary to have one {\em free} measurement choice. 
For example, it can suffice for Bob to make only a single generalized measurement~\cite{Saunders12}. An alternative scheme, whereby Alice uses the same randomly generated setting as Bob, but can find it out only after Bob's photon 
is in his `lab' (his trusted space), was implemented in Ref.~\cite{Wittmann12}. In that experiment, Alice's and Bob's outcomes were space-like separated, as required by the {\em locality} loophole, and it also closed the efficiency loophole (three years before the first LHF Bell experiments~\cite{LHFnature,LHFprl1,LHFprl2}).  {Proving that an EPR-steering experiment is not affected by any potential {\em memory} loophole has never been done, as it would require the development of an entirely new and nontrivial analysis method. In the current work we close the efficiency, freedom-of-choice, and locality loopholes, but we analyze our data in a way that assumes that experimental trials are independent.}

\begin{figure}[tb]
	\begin{center}
		\includegraphics[width=\columnwidth]{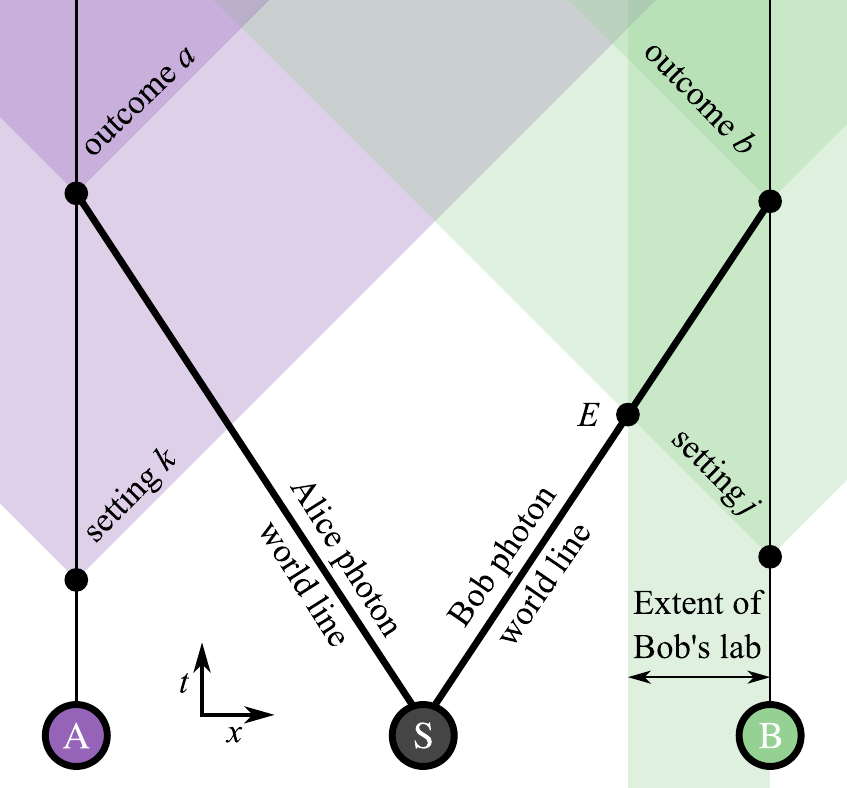}
		\caption{Minkowski diagram illustrating the relations between spacetime events (black dots) at Alice (A), Bob (B), and the source (S) that are needed to close the locality loophole, for a simple example in which all parties are arranged collinearly. A trial begins when the source emits an entangled pair of photons. If these photons travel in fiber, they will travel at roughly 2/3 light speed, as indicated by the slope of their world-lines. While the photons are in flight, measurement settings $k$ and $j$ are randomly chosen, defining their associated future light cones. The event $E$ is the point in spacetime where the world line of Bob's photon intersects the light cone corresponding his setting choice $j$. Bob's `lab' must be defined to include the event $E$, and the event $E$ must lie outside of the future light cone of Alice's setting $k$. Finally, Alice must complete her measurement and record her outcome $a$ outside of the future light cone of Bob's outcome $b$.}\label{fg:timings}
	\end{center}
\end{figure}

In our current experiment, we use measurement choice on both sides, and so the LHF nature is 
ensured in a similar way to the bipartite photonic LHF Bell experiments~\cite{LHFprl1,LHFprl2}, with two random number generators and fast electro-optic switching. The main difference is that, as in Ref.~\cite{Wittmann12}, Bob has a lab, as illustrated in Fig.~\ref{fg:timings}, in which he trusts the operation of his devices and his quantum mechanical description of them. This lab is taken to be large enough that it contains the event $E$ defined as the intersection of the future light-cone emanating from the event $j$ of Bob's random number generation with the world-line of the photon which Bob will ultimately detect. Note this world-line only needs to be defined inside Bob's lab (where it is indeed defined because Bob trusts his apparatus) in order to define $E$. While $E$ is a point in space-time, not a point in space, it is natural to define Bob's lab to be stationary in the rest frame of his apparatus, so that $E$ effectively determines its extent. As in LHF Bell experiments, we must trust that Alice's and Bob's random number generators make their basis choices at the times and locations we set them to. Finally, event $E$ must not be in the future light cone of Alice's random number generation $k$, while the event $b$ of Bob's measurement result becoming macroscopic (and thus, we'll assume, potentially `visible' to an adversary) must not be in the past light cone of the event $a$ of Alice reporting her result.
 
The reason Bob needs a trusted space which includes event $E$ satisfying the above conditions is as follows. In a steering scenario, Alice is not trusted, and she could have an accomplice, Fenella, who can act on her behalf anywhere except in Bob's trusted space. If Fenella were present at event $E$, or any subsequent point on the world-line of the photon going to Bob's detectors, then she could make the state of that photon depend on Bob's measurement choice $j$. This violates one of the fundamental assumptions behind EPR-steering inequalities, that $j$ is uncorrelated with Bob's LHS. It would easily allow Alice to cheat as follows: Knowing the setting $j$, Fenella could prepare the state so that Bob always gets the result $+1$, and thus a steering inequality of the form we later consider could attain its logically maximum value (and therefore violate the EPR-steering bound) if Alice simply always says $+1$ too. This means that $E$ has to be off-limits to Fenella. In other words, it has to be in Bob's trusted space. Note that this is in contrast to Bell inequalities, where there is no LHS assumption, and no trusted space. 

The event $E$ cannot be in the future light-cone of the generation of Alice's setting $k$ because if it were then Fenella could make the state Bob is to detect depend on $k$, which again violates the fundamental assumption of EPR steering. In this case, it is analogous to the Bell condition that the event of Bob's outcome $b$ cannot occur in the future light-cone of the generation of Alice's setting $k$. Similarly, an untrusted Alice cannot be allowed to delay the generation of her outcome $a$ until it is in the future light-cone of Bob's result $b$ appearing, which in this case is exactly the same in EPR steering as it is in the Bell case\footnote{Both of these conditions (that involving $E$ and $k$, and that involving $a$ and $b$) are automatically guaranteed by the conditions that are used in a LHF Bell scenario, namely that  each party's choice of setting is space-like separated from the other party's observation. However, we note that these LHF-Bell conditions are not necessary for a LHF test of EPR-steering correlations.}. 
Finally, note that Bob's detectors being in his trusted space means that he is allowed to post-select on a successful photon-detection. 
 
The extra feature of the experiment reported here is that we allow, in the theory, Alice to send one bit of information FTL to her accomplice Fenella. This means that Fenella can make, in any trial, Bob's photon have one of two possible polarization states after she knows Alice's setting $k$ but before it enters Bob's lab. That the information would have to be sent FTL is guaranteed by the fact that event $E$ (at which is present the photon Bob will detect) is space-like separated from Alice's random number generation $k$, so any event that can influence the state of the photon at $E$ must also be space-like separated from $k$. Because it rules out one bit of FTL communication as an explanation for EPR steering, our experiment is stronger, and considerably more difficult to achieve, than that of Ref.~\cite{Wittmann12}. 
Finally, we note that it would not be possible to do an experiment similar to the current one that rules out more than one bit of FTL communication in a Bell scenario, because---under the usual Bell assumptions in which both Alice and Bob are untrusted---one bit of communication suffices to simulate any correlations from projective measurements on a singlet state~\cite{Toner03}.

\section{EPR-steering inequality with bounded one-way communication}\label{sc:newInequality}

We consider an EPR-steering task with bounded one-way communication. That is, the model to disprove is one with a LHS for Bob, but where Alice can send a classical message, taking one of $d$ possible values, to Bob's side to affect his state before his detection. Here, $d$ is an integer between $1$ (no message) and $n-1$, where $n$ is the number of settings Alice uses. There is no point considering $d \geq n$ because that would allow perfect FTL communication of Alice's setting $k$ to Bob's side---technically, to event $E$---which would allow perfect correlations even with the LHS assumption, and so could never be disproven experimentally. It is convenient to define $d=2^{H_0}$, where $H_0$ is the number of bits of communication in a max-entropy sense. Recently, Refs.~\cite{sr_Nagy,pra_Sainz} studied the quantification of EPR steering according to the classical message size $c$ (i.e. $H_0$) needed to simulate an assemblage without entanglement. Here an assemblage is a complete description of Alice's ability to steer Bob in a given experiment --- the set of all Bob's states to which she can steer, and associated probabilities, indexed by her settings and outcomes. In Ref.~\cite{sr_Nagy} the set of assemblages which have an LHS model when $c$ bits of communication are sent from Alice to Bob was shown to have an efficient semidefinite program (SDP) formulation. They proved that infinite communication is necessary to simulate the maximally entangled state. On the other hand, in Ref.~\cite{pra_Sainz} the LHS-robustness was shown to provide an upper bound on the amount of communication, and infinite communication cost was shown to be necessary even for some impure states that were not full-rank. 

In this paper we adopt the approach of generalizing the linear steering inequalities of Refs.~\cite{DJS10,Bennet12,pra_Evans}. Bob is assumed to have some LHS $\rho$, and---under this assumption---a violation indicates that Alice's measurement choice must have an effect on this state. Our formulation establishes a correlation bound,  {$h^n_{H_0}$}, beyond which 
one would require $H_0$ bits of superluminal communication to 
replicate the effect of Alice's choice, in a LHS model. 
The bound  is determined by optimizing the following LHS (no-entanglement) protocol. (1) Bob sets his apparatus to measure the observable $\hat B_j$ with randomly chosen setting index $j\in\{1,\ldots, n\}$. {(2) At the same time (i.e. as a space-like separated event), Alice generates an outcome $a_k$ (from a randomly chosen setting  $k\in\{1,\ldots, n\}$). (3) Alice sends a{n} FTL message $l\in\{1,\ldots,d\}$ to her accomplice, Fenella, located near Bob's laboratory. (4) Knowing $l$, Fenella generates a state $\ket{l}_{\rm  F}$, one of $d$ possible states $\{\ket{l}_{\rm  F}\}_{l=1}^{d}$ she could prepare, and feeds it into Bob's laboratory. Note that the states $\{\ket{l}_{\rm  F}\}_{l=1}^{d}$ need not be mutually orthogonal, even for $d=2$. 

Two-outcome measurements yield a binary variable, $+1$ or $-1$. Physically, in our experiment, this corresponds to a click in one of two detectors, at each side\footnote{It is possible for \textit{both} detectors on either side to click in the same trial. We discuss how we treat those events in Section~\ref{sc:results}.}. Due to the limited transmission and detection efficiency of the photons, a third outcome $0$ has to be considered to represent no detection. According to the LHS model, Bob trusts his apparatus, including his two detectors, so he can discard those experimental trials where he fails to detect a photon. That is, Bob can make the fair-sampling assumption without opening the efficiency loophole, and describe his output by the binary variable $b_j \in \{+1,-1\}$. However, because Alice is not trusted, she is not allowed to discard any results, and her output is a ternary variable $a_k\in\{+1, 0, -1\}$. After all measurement trials are complete, Alice sends her data to Bob. Bob discards all data for trials in which he did not detect an outcome, and then he computes an appropriate steering correlation function $S_{n}$ and checks whether the inequality  
\begin{equation}
S_{n} \leq {h^{n}_{H_0}} \label{ginequality}
\end{equation}
is violated for the appropriately calculated bound $h^{n}_{H_0}$.  If so, they have demonstrated EPR steering requiring more than $H_0$ bits of FTL communication. Note that the dependence on $n$ here is shorthand for dependence on the set of $n$ observables $\{\hat B_j\}_{j=1}^n$ that Bob performs. Neither the correlation function nor the bound make any assumptions on the nature of the measurements that Alice performs, apart from their having outcomes $a_k\in\{+1, 0, -1\}$.

The task, then, is to choose a suitable steering correlation function, or family of correlation functions, and to compute the bound that can be achieved with a{n} LHS model supplemented by allowing Alice $H_0$ bits of FTL communication. We design the steering correlation functions specifically for our experiment, in which the two-photon state shared by Alice and Bob is close to a maximally entangled singlet state $\ket{\Psi^-}$ that has been subjected to loss, and Alice's $n$ measurement axes are intended to be the same as Bob's. For simplicity, we number the corresponding pairs of axes the same, and label the directions corresponding to the results ($\pm1$) in opposite directions. Thus, Alice's and Bob's results will be maximally correlated if they choose the same setting. Such correlations are best for demonstrating steering~\cite{DJS10}. Also, for all $j$ and $k$ the mean of $b_j$ and $a_k$ will be close to zero, and $|b_j|$ is identically $1$ (because Bob can post-select).
Finally, we aim to implement {a set of} measurements {$\{\hat{B}_j\}$ with a symmetry property that ensures} that no measurement is special. Specifically, we choose a set of measurements for which, given any two measurements, there exists a rotation of the axes that can take the first to the second while leaving Bob's total set of measurement directions unchanged. {That is, 
$\forall j, j' \, \exists \hat U \, : \, \hat U^\dagger \hat B_j \hat U =  \hat B_{j'} \textrm{ and } \{\hat U^\dagger \hat B_i \hat U\}_i = \{\hat B_i\}_i.$ (As a concrete example, the set of measurements that contains the three Pauli observables $\{ \hat{\sigma}_x,\hat{\sigma}_y,\hat{\sigma}_z\}$ satisfies this symmetry requirement.)}

Based on these considerations we can impose, with almost no loss of effectiveness, simplifying symmetries on the form of our correlation function:  invariance under interchanging the results $\pm 1$ for $a$ and $b$ jointly; and invariance 
under permutation of setting values $j$ and $k$ jointly. Imposing these symmetries means that the function weighs all correlations equally, and thus we obtain a family of steering correlation functions with only one real parameter, $r_{H_0}$:
\beq
S_{n}~({r_{H_0}})=\frac{1}{n}  \sum_{j,~k=1}^{n}\delta_{j,k}\left(\langle a_{k} \hat{B}_{j}\rangle-{r_{H_0}} 
 \langle|a_{k}|\rangle\right).
 \label{eq:genCorrFunction}
\eeq
In Eq.~\ref{eq:genCorrFunction}, Alice's result $a_k$ is to be considered a random variable, and so is Bob's result $b_j$, but we represent 
the latter by its associated Hermitian operator $\hat B_j$ to emphasize that (unlike in a Bell correlation) we trust this 
description of Bob's measurement process. That is, deriving the EPR-steering bound on this correlation (below) makes 
direct use of the operator properties of the $\hat B_j$.
Note that the use of the above symmetries in motivating the above correlation function is for simplicity only;
the validity of the inequality we will ultimately derive does not depend on assuming that these symmetries in the state and measurement settings are satisfied.  

Once the parameter {$r_{H_0}$} is chosen, the bound {$h^n_{H_0}$} can then be calculated via 
\beq \label{eq:maxs}
{h^n_{H_0}} =  \max_{\alpha,\,\ell,\,\{\ket{l}_{\rm F}\}_{l=1}^d} \frac{1}{n}\sum_{j=1}^{n} \delta_{a,\alpha(j)} \left(-{r_{H_0}} |a| + \sum_{l=1}^{d} \delta_{l,\ell(j)}\, a\, {}\bra{l}_{\rm F} \hat{B}_{j}\ket{l}_{\rm F} \right) ,
\eeq
where $\ell$ indicates a deterministic function{\footnote{{In principle, Alice and Fenella could update this function based on the results of past measurement trials. However, since we assume that all trials are independent and identical, we do not consider that here.}}} from $\{1, \dots, n\}$ to $\{1, \dots ,d\}$ (recall $d=2^{H_0}$), which specifies the strategy for Fenella to prepare the states $\{|l\rangle_{\rm  F}\}_{l=1}^d$ related to Alice's measurement setting $k$, and $\alpha$ indicates a deterministic function from $\{1, \dots, n\}$ to $\{+1, 0, -1\}$, which determines Alice's outcomes. Note that the linearity of the correlation function with respect to Bob's observables means that it can always be maximized by a deterministic strategy, allowing us to calculate the bound in this way. 

We define Alice's heralding efficiency through the method outlined by Klyshko~\cite{Klyshko}. That is, it is the probability with which a detection by Bob heralds a detection by Alice. It is given by the expression $\frac{1}{n}\sum_{k=1}^{n}\langle |a_{k}|\rangle$, and in the below, we will simply call this Alice's efficiency, $\eta_A$. The purpose of $r_{H_0}$ is to make the steering demonstration as loss-tolerant as possible. That is, to allow violation for a value of Alice's heralding efficiency which is as small as possible given the other parameters in the experiment (the degree of mixture in the lossy maximally entangled singlet state, the value of $H_0$, and the measurement axes).

\begin{figure}[b]
	\begin{center}
		\includegraphics[width=\columnwidth]{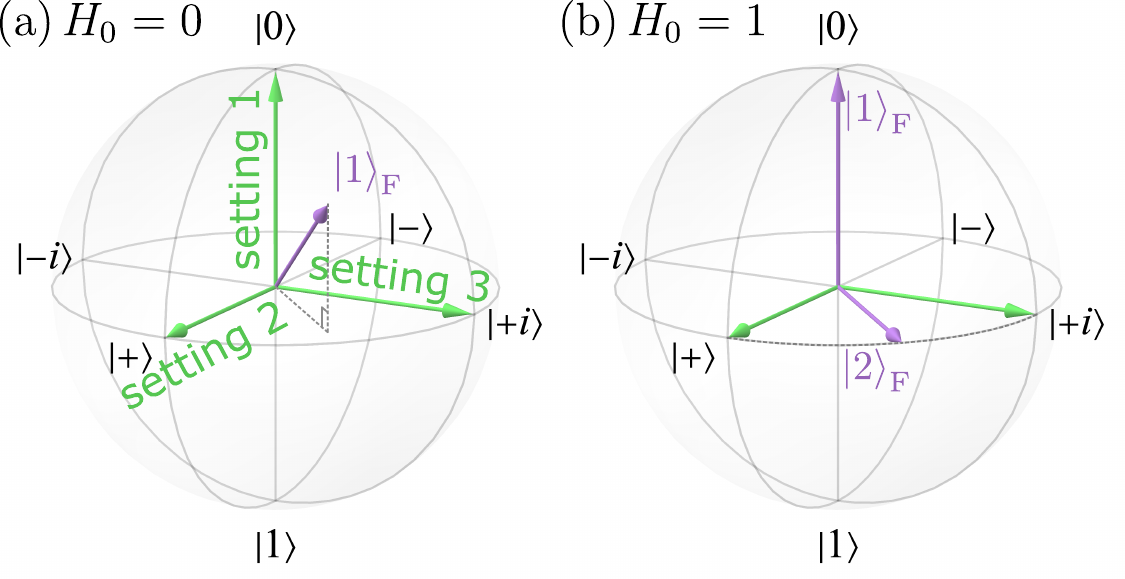}
		\caption{Sketch of Bob's three ideal measurement directions (green arrows represent the +1 eigenstate of each measurement setting) and Alice's optimized cheating ensembles generated by Fenella  $\{|l\rangle_{\rm F}\}$ (purple arrows) in the Bloch spheres. (a) Without any information transferred ($H_0=0$), there is only one optimized cheating state $|1\rangle_{\rm F}$ for all three measurement directions; (b) When $H_0=1$ bit messages are transferred, there are many possible optimal cheating strategies. One such strategy is as follows: if Alice measures in setting $k=1$, she sends Fenella the message $l=1$ and Fenella prepares the state $\ket{1}_\mathrm{F}$. Otherwise, Alice sends the message $l=2$ and Fenella prepares the state $\ket{2}_\mathrm{F}$.}\label{ideal}
	\end{center}
\end{figure}

\section{Inequalities for octahedral setting arrangement} \label{sec:Ineqoct}

In our experiment, we use $n=3$, the minimum number required to rule out LHS models with up to 1 bit of FTL communication}. We use an octahedral arrangement for the measurement axes. That is, the labels $\{1,2,3\}$ correspond to states $\ket{0}$, $\ket{+}$, and $\ket{+i}$, as illustrated in Fig.~\ref{ideal}. 
This simplifies the derivation of the inequality, allowing us to obtain analytical bounds, which we present in this section. In a later section, where we allow for systematic and statistical errors in Bob's settings, we obtain the bounds numerically. 
In both cases we find the bounds by an exhaustive search over deterministic strategies. 
We begin by considering the no message ($H_0=0$) case, for which optimising {$r_{H_0}$} gives the same loss-tolerant inequalities previously derived for EPR-steering with no FTL message assistance~\cite{Bennet12,pra_Evans}.

\subsection{Derivation of the inequality for the no-message case}

If Alice, the untrusted party, had perfect detectors, she could declare non-null results for every trial and for all $n$ settings, so that we would have $\langle |a_{k}|\rangle=1$, and we do not need the parameter {$r_0$}. When there is no message transferred ($H_0=0$), Fenella generates the same state $\{|l\rangle_{\rm F}\}$ ($l=1$) for all three settings. The optimal orientation is towards the face-center of the octahedron used to define three measurement axes, as delineated in Fig.~\ref{ideal}(a). This leads to the following EPR-steering inequality
\begin{equation}
\frac{1}{3} \sum_{j,~k=1}^{3} \delta_{j,k} \langle a_{k} \hat{B}_{j}\rangle \leq \frac{\sqrt{3}}{3},\label{s30eta1}
\end{equation}
The bound ${h^{3}_{0}}=\sqrt{3}/3$ coincides with the inequality previously given in Ref.~\cite{DJS10}.

If Alice sometimes reports null results, the parameter {$r_0$} is now relevant, 
and a new search for the optimal solutions for $\alpha(j)$ and $\{\ket{l}_{\rm F}\}$ 
must be undertaken for each value of {$r_0$} in \erf{eq:maxs}
(Note that in this no-message case, the function $\ell$ is trivial, so no search 
over this is needed). In addition, we wish to choose {$r_0$}  optimally for a given value of 
$\eta_A$, Alice's efficiency. Note that while we assume, 
for the purpose of optimizing $r_0$, that Alice's 
efficiency is independent of $k$ ($\eta_A =\frac{1}{n}\sum_{k=1}^{n}\langle |a_{k}|\rangle$), and that Alice's two detectors have the same efficiency, 
these assumptions are not required for the inequality itself to be valid. 
An optimal {$r_0$}  for a given $\eta_A$ is the one which allows the inequality to be violated most easily with our model of the state, a loss-depleted Werner state \cite{werner}
\beq
W_\mu =\mu \ket{\Psi^-}\bra{\Psi^-} +(1-\mu)\textbf{I}/4
\eeq
with singlet-proportion $\mu\in[0,1]$. For such a model, the residual between the left-hand-side and right-hand-side of \erf{ginequality} is given by 
\beq
{R^3_{0}} = \eta_A(\mu - r) - {h^3_0}(r). \label{res0}
\eeq

Maximizing this residual for any $\eta_A$ (which is equivalent to minimizing the $\mu$ at which a 
positive residual is possible), we find that the optimal {$r_0$}, and the corresponding bound, is piecewise constant, changing at the points $\eta_A = 1/3$ and $2/3$. Examining these points analytically, and using the $\eta_A=1$ analysis above, we are able to find the relevant values of {$r_0$}, and the bound, analytically. We show how this is done for one example in Appendix A. We find 
\begin{eqnarray}
{r_0}&=&
\begin{cases}
\sqrt{2}-1  &\hbox{if $\frac{1}{3} \leq\eta_A< \frac{2}{3}$}, \\
\sqrt{3}-\sqrt{2} &\hbox{if $\frac{2}{3} \leq\eta_A< 1$}, \\
\end{cases}\nonumber\\
{h^3_0}&=&
\begin{cases}
\frac{2-\sqrt{2}}{3} & \hbox{~~if $\frac{1}{3} \leq \eta_A < \frac{2}{3}$},\\
\frac{3\sqrt{2}-2\sqrt{3}}{3} & \hbox{~~if $\frac{2}{3} \leq\eta_A < 1$}. \\
\end{cases}\label{ideal0}
\end{eqnarray}
If Alice reports too many null results, so that $\eta_A<1/3$, the optimal gain factor is ${r_0}=1$, 
and ${h^{3}_{0}}= 0$. This means that the inequality is impossible to violate. This is because, below this detection-efficiency threshold, the steering correlation function $S_3$ cannot be positive even with the maximally entangled states~\cite{Bennet12}.

\subsection{Derivation of the inequality when one FTL bit is transferred}

Now allowing Alice the assistance of an $H_0=1$ bit message, we must include the function $\ell: \{1,2,3\} \to \{1,2\}$ which allows Fenella to choose a state from $\{|l\rangle_{\rm F}\}_{l=1}^2$ given Alice's measurement setting {$k$}. If Alice declares a non-null result in every round, then we find the inequality 
\begin{equation}
\frac{1}{3} \sum_{j,k=1}^{3}\delta_{j,k} \langle a_{k} \hat{B}_{j}\rangle \leq\frac{1+\sqrt{2}}{3} ,\label{s31eta1}
\end{equation} 
with the optimized parameter ${r_1}=0$. Here, the bound ${h^{3}_{1}}=(1+\sqrt{2})/3$ can be achieved by appropriately choosing $\{\ket{l}_{\rm F}\}_{l=1}^2$, $\alpha(j)$ and $\ell(j)$. For instance, the optimal strategy  for Alice/Fenella with $\alpha(j)=+1$ for all $j$, and $\ell(1)=1$, $\ell(2)=\ell(3)=2$ is shown in Fig.~\ref{ideal}(b). 

Turning now to the case of where Alice has to deal with loss (or is trying to cheat by declaring 
null results), the maximization has to be done including the parameter {$r_1$}, and {$r_1$} has to be chosen optimally. Again, we find the solution analytically: 
\begin{equation}
{r_1}=\sqrt{2}-1,~{h^{3}_{1}}=\frac{4-2\sqrt{2}}{3}~~~\text{if}~\frac{2}{3} \leq\eta_A < 1.\label{ideal1}
\end{equation} 
Now, when Alice's efficiency is less than $2/3$, the optimal ${r_1}=1$ and ${h^{3}_{1}}=0$, leading to a trivial (unable to be violated) inequality. This is as expected, as the assistance of transferred messages can make it easier for a lossy Alice to find a cheating strategy, which in turn means more stringent conditions for her efficiency to demonstrate EPR-steering. The residual from \erf{ginequality}, 
\beq
{R^3_{1}} = \eta_A(\mu - r) - {h^3_1}(r), \label{res1}
\eeq
shows that to have the ${R^n_{1}}>0$, the required purity $\mu \to 1$ as $\eta_A$ approaches $2/3$ from above. The details of the optimization procedure leading to Eqs.~(\ref{s31eta1}) and (\ref{ideal1}) can be found in Appendix A.
When the measurements actually implemented by Bob are different from the above ideal case, we have to reoptimize the parameters {$r_{H_0}$} to obtain the corresponding bounds; this will be covered in more detail in Sec. \ref{realbound}.

\section{Experimental methods}\label{sc:experiment}

We perform a test of our new EPR-steering inequality with polarization-entangled photon pairs that are distributed to two distant measurement stations we name Alice and Bob.

\begin{figure}
	\begin{centering}
		\includegraphics[width=\columnwidth]{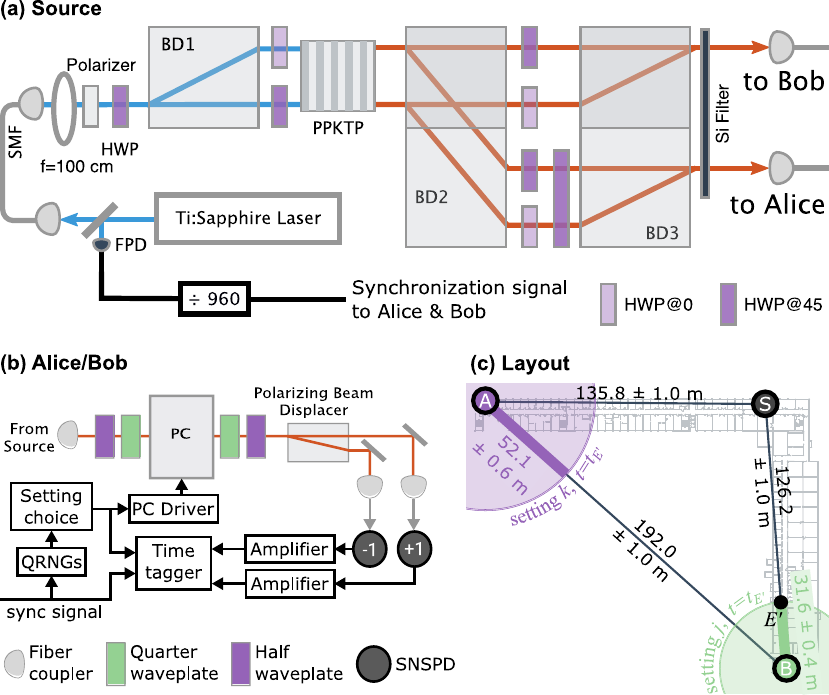}
		\caption{Experimental test of EPR steering with one bit of FTL communication. (a) The source produces polarization-encoded photon pairs in the maximally entangled singlet state via type-II spontaneous parametric downconversion. (b) Alice's and Bob's labs where the photons' polarizations are measured; a Pockels cell allows fast switching between three mutually unbiased measurement bases. (c) Layout of the experiment, showing the relative locations of Alice (A), Bob (B) and the source (S), overlaid on a floor plan of the section of the building where the experiment took place. The spacetime event $E'$ (black dot) is our conservative estimate of the event $E$ (see Sec.~\ref{sc:experimentalLocalityConditions} for details). At the time of $E'$, $t_{E'}$, the region of space inside the future light cone of Bob's setting choice $j$ is represented by the green circle, and the region of space inside the future light cone of Alice's setting choice $k$ is represented by the purple circle. The faded rings outside the green and purple circles represent the uncertainties on their radii. We define Bob's lab (in which he trusts his characterization of his equipment) to contain his measurement apparatus plus the final {$31.6\pm 0.4$ m} of fiber leading to his measurement apparatus. FPD: fast photodiode; SMF: single-mode fiber; BD: polarizing beam displacer; PPKTP: periodically poled potassium titanyl phosphate; {QRNG: quantum random number generator;} PC: Pockels cell; SNSPD: superconducting nanowire single-photon detector.  \label{fg:setup}}
	\end{centering}
\end{figure}

The entangled pair source~\cite{LHFprl2} is displayed in Fig.~\ref{fg:setup}(a). A Ti:Sapphire laser emits a beam of pump pulses with wavelength 775 nm at a repetition rate of 80 MHz. A small fraction of the beam is picked off and sent to a fast photodiode (FPD) which sends an electronic synchronization pulse to both Alice and Bob once every 960 laser pulses. The remainder of the pump beam is sent through a single-mode fiber to clean up its spatial mode. After exiting the fiber, the beam is gently focused and prepared in the diagonal polarization state before it is equally split into two paths by a polarizing beam displacer (BD1). In each path, photon pairs are produced via type-II spontaneous parametric downconversion in a 20 mm-long periodically poled potassium titanyl phosphate crystal placed at the pump beam's focus. Polarizing beam displacers BD2 and BD3 and a series of half waveplates are used to recombine the beams into two paths (with one photon in each path). A silicon window filters out the pump beam, then each photon is coupled into single-mode  fiber over which it is sent to Alice's or Bob's measurement station. By controlling the polarization rotation that occurs inside each fiber, we ensure that the pair arriving at Alice's and Bob's measurement stations is in the maximally entangled singlet state $\frac{1}{\sqrt{2}}(|H\rangle|V\rangle - |V\rangle|H\rangle)$, where $|H\rangle$ ($|V\rangle$) denotes horizontal (vertical) polarization.

The measurement stations at Alice and Bob are nominally identical, and one of them is represented in Fig.~\ref{fg:setup}(b). We perform measurement trials at a rate of roughly 83 kHz -- each electronic synchronization pulse from the source triggers a new measurement trial. At the beginning of each trial, a pair of quantum random number generators{~\cite{abellan15,wayne18}} produces a pair of random bits to determine the setting for that trial. If these two output bits are 00, 01, or 10, we measure in setting 1, 2, or 3 respectively, and the setting choice determines which voltage is applied to a Pockels cell. (One fourth of the time the two output bits will be 11; we ignore these trials since we want each of the three settings to occur with equal probability.) When a photon arrives at a measurement station, it is coupled into free space and transmitted through four waveplates and a Pockels cell before its $|H\rangle$ and $|V\rangle$ components are separated with a polarizing beam displacer. Each output of the beam displacer is then coupled into single-mode fiber where it is sent to one of two superconducting nanowire single-photon detectors (SNSPDs), labelled\footnote{In Fig.~\ref{fg:setup}(b) we denote the labelling for Bob's measurement apparatus, where detections in the transmitted output port of the BD are labelled `$+1$' and detections in the displaced output port are labelled `$-1$'. In Alice's measurement apparatus, the labels are swapped.} with the outcomes `+1' and `-1'. Output pulses from the SNSPDs are amplified before being recorded by a time tagger. We map the computational basis $\{|0\rangle,|1\rangle\}$ to polarization states with the assignments $|0\rangle \rightarrow |V\rangle$ and $|1\rangle \rightarrow |H\rangle$. The waveplate angles and Pockels cell voltages are set to (ideally) 
realize the measurements in Fig.~\ref{ideal}. That is, ideally, setting 1 corresponds to a measurement in the {$\{|0\rangle,|1\rangle\}$ } {(}$\{|V\rangle,|H\rangle\}${)} basis, setting 2 to the {$\{|{+}\rangle,|{-}\rangle\}$ } {(}$\{|D\rangle,|A\rangle\}${)} basis, and setting 3 to {$\{|{+i}\rangle,|{-i}\rangle\}$ } {(}$\{|L\rangle,|R\rangle\}${)}.
Here, $|D\rangle$, $|A\rangle$, $|L\rangle$, and $|R\rangle$ correspond to the diagonal, antidiagonal, left-circular, and right-circular polarization states. We provide full details of how we determine which waveplate angles and Pockels cell voltages to use in Appendix~\ref{ap:pockelscell}.

\subsection{Locality conditions}\label{sc:experimentalLocalityConditions}

In our experiment, the pair source and Alice's and Bob's measurement stations are arranged in a triangular configuration (Fig.~\ref{fg:setup}(c)). Before assessing if the locality conditions are met, we must first define the extent of Bob's lab, i.e. the size of the area in which Bob trusts his characterization of his equipment, which is defined by the spacetime event $E$. {Recall that the spacetime event $E$ is defined as the point in space-time at which the light cone emanating from Bob's random number generators (and potentially carrying information about Bob's measurement setting choice) intersects with the world line of the photon travelling to Bob's lab (Fig.~\ref{fg:timings}). One piece of information we need to find $E$ is the exact path that Bob's photon takes while it is inside his lab. Because Bob's photon travels in fiber along the hallways and through the ceiling of the building we run the experiment in, we only approximately know where the fiber lies. 
We do, however, have an accurate characterization of the position of the Pockels cell in Bob's measurement setup, as well as the time that Bob's photon arrives at his Pockels cell. Therefore, we make the conservative assumption that, inside his lab, Bob's photon travels in a straight line in fiber between the source and his Pockels cell, which is shorter than the true path that Bob's photon actually takes. We then calculate the position of the spacetime point of the event $E'$ (Fig.~\ref{fg:setup}(c)), which would be equivalent to $E$ if this assumption about the fiber inside Bob's lab were true, and we define Bob's lab to include the point $E'$. Defining Bob's lab in this way gives us a conservative \emph{overestimate} of the size of Bob's lab (and thus ensures it contains the point $E$).}  We display $E'$ and Bob's setting light cone at time $t_{E'}$ (i.e. the time of event $E'$) in Fig.~\ref{fg:setup}(c).  The point $E'$ is {$31.6 \pm 0.4$ m} from Bob's random number generators, so we define the trusted equipment in Bob's lab to include that final length of fiber.

To close the locality loophole, there are two conditions that need to be met. First, a light-speed signal carrying information about Alice's measurement setting $k$ must not be able to reach Bob's lab before the time of the event $E$. Our definition of $E'$ guarantees that $E'$ occurs later in time than $E$. In addition, $E'$ is closer to Alice's lab than any part of the fiber inside Bob's lab, and this guarantees that the distance between Alice's lab and $E'$ is shorter than the distance between Alice's lab and $E$. As a result of these two facts, any signal carrying information about $k$ needs to travel faster to influence $E$ than it does to influence $E'$, and this implies that if $k$ and $E'$ are space-like separated then  $k$ and $E$ must be space-like separated as well. Alice's first random number generator fires $230\pm2$ ns before time $t_{E'}$, and we consider this the earliest moment that a light speed signal containing information about Alice's setting choice could begin propagating from her lab. Hence, at time $t_{E'}$, the light cone with information about Alice's setting $k$ has traveled {at most} {$52.1 \pm0.6$ m} from Alice's random number generators---and has not reached Bob's lab---ensuring that this first locality condition is satisfied (Fig.~\ref{fg:setup}(c)). 

The second condition that needs to be satisfied is that Alice's measurement outcome $a$ must be recorded before the light cone carrying information about Bob's outcome $b$ reaches Alice. We consider Alice's measurement complete (and her guess of Bob's outcome recorded) the moment the amplified electronic pulse from either of her detectors reaches her time tagger. We define the earliest possible time that Bob's measurement could be considered complete as the moment that his photon impinges on one of his SNSPDs. With these definitions, Alice records her measurement outcome at least $44.8 \pm1.3$ ns before Bob completes his measurement, ensuring that the second locality condition is satisfied.

\subsection{Tomography on Bob's measurements and relevant bounds}\label{realbound}

\begin{figure}[b]
	\begin{center}
		\includegraphics[width=\columnwidth]{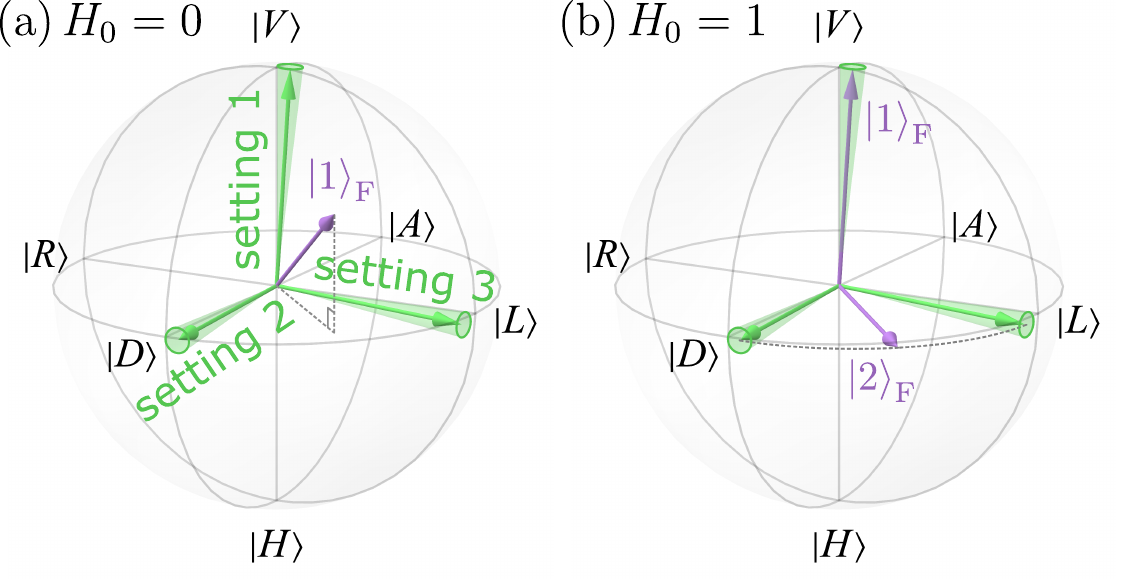}
		\caption{Characterization of Bob's three actual measurement directions (green arrows) and Alice's optimized cheating ensembles generated by Fenella $\{|l\rangle_{\rm F}\}$ (purple arrows) in the Bloch sphere. The green arrows represent our best estimate of Bob's actual measurement settings and with five-sigma certainty, the actual Bloch vectors representing each measurement setting lie inside the corresponding cones of uncertainty.  (a) Alice's optimal cheating state  $|1\rangle_{\rm F}$ if no information is transferred ($H_0=0$); (b) Alice's optimal cheating states $|1\rangle_{\rm F}$ and $|2\rangle_{\rm F}$ if an $H_0=1$ bit message is transferred. Details of how we calculate Alice's optimal cheating ensembles are summarized in Appendix~\ref{sc:actualMeasTheory}.}\label{actual}
	\end{center}
\end{figure}

To obtain rigorous EPR-steering inequalities we need to take into account that the measurements actually implemented by Bob are different from the ideal ones assumed for the analytical bounds obtained above. To account for this, we perform a parametric bootstrapping measurement tomography routine which provides us with estimates of the positive-operator valued measures (POVMs) representing each of the measurement settings implemented by Bob. To be conservative, we constrain our estimates of Bob's measurements to be both noiseless and projective (since this is the optimal scenario for a dishonest Alice\footnote{Alice and Fenella's strategy of controlling the state that is sent to Bob in order to influence his measurement outcomes will work best if Bob's measurements are both noiseless and projective. Adding noise to Bob's measurements only decreases how accurately his outcomes can be predicted, which in turn decreases Alice's and Fenella's probability of success~\cite{Smith12}}.), meaning each POVM can be represented by the Bloch vector that represents the eigenstate corresponding to the +1 eigenvalue of the measurement operator. Our tomography method also provides estimates of the uncertainties on each of these Bloch vectors, in the form of an angular uncertainty on the direction that they point. We provide our best estimates of the Bloch vector representing each of Bob's measurement settings, as well as their uncertainties, in Table~\ref{table1}. We illustrate these Bloch vectors and their uncertainties with {cones} in Fig.~\ref{actual}; the region enclosed by each cylinder represents the region that we expect---with five-sigma certainty---the actual Bloch vector representing that measurement setting to lie in. Full details of our measurement tomography technique are provided in Appendix~\ref{sc:BobMeas}. 

\begin{table}[!htb]
\centering 
	\renewcommand\arraystretch{1.25}
	\caption{Results of tomography on Bob's measurements.}\label{table1}
	\begin{tabular*}{\columnwidth}{ c@{\extracolsep{\fill}} S[table-format=2.4] S[table-format=2.4] S[table-format=2.4] }
\hline\hline\Tstrut
{}\Tstrut & {Setting 1}& {Setting 2}& {Setting 3}\\
\hline
		X\Tstrut & -0.0502 & 0.9984 & 0.1019  \\
		Y\Tstrut & 0.0419 & 0.0559 & 0.9944  \\
		Z\Tstrut & 0.9978 & -0.0089 & -0.0276 \\
		    	Angular uncertainty, $\sigma_j$ (rad)\Bstrut & 0.0114 & 0.0114 &0.0114 \\
\hline\hline
\end{tabular*}
\end{table}

Using the above tomographic results, we recalculate Alice's optimal cheating ensembles based on Bob's real measurements, as delineated in Fig.~\ref{actual} (a) and (b). For a completely rigorous EPR-steering test we must also include uncertainties in our characterization of Bob's actual settings, at the five-sigma level; see Appendix~\ref{sc:actualMeasTheory} for more details. By using high-efficiency detectors, Alice's average heralding efficiency over all three settings --- as defined at the end of Section~\ref{sc:newInequality} to be $\eta_A = \frac{1}{3}\sum_{k=1}^{3}\langle |a_{k}|\rangle$ --- is measured to be {$(74.8\pm0.1)\%$}. For this efficiency level and for Bob's measurements (including errors), the gain factor {$r_{H_0}$} and the bound {$h^n_{H_0}$} are reoptimized, yielding
\begin{eqnarray}
{r_{0}}&=&0.4046,~{h^{3}_{0}}=0.2548,\nonumber\\
{r_{1}}&=&0.5930,~{h^{3}_{1}}=0.2713.\label{actrh}
\end{eqnarray}
\begin{figure}[t]
	\begin{center}
		\includegraphics[width=0.85\columnwidth]{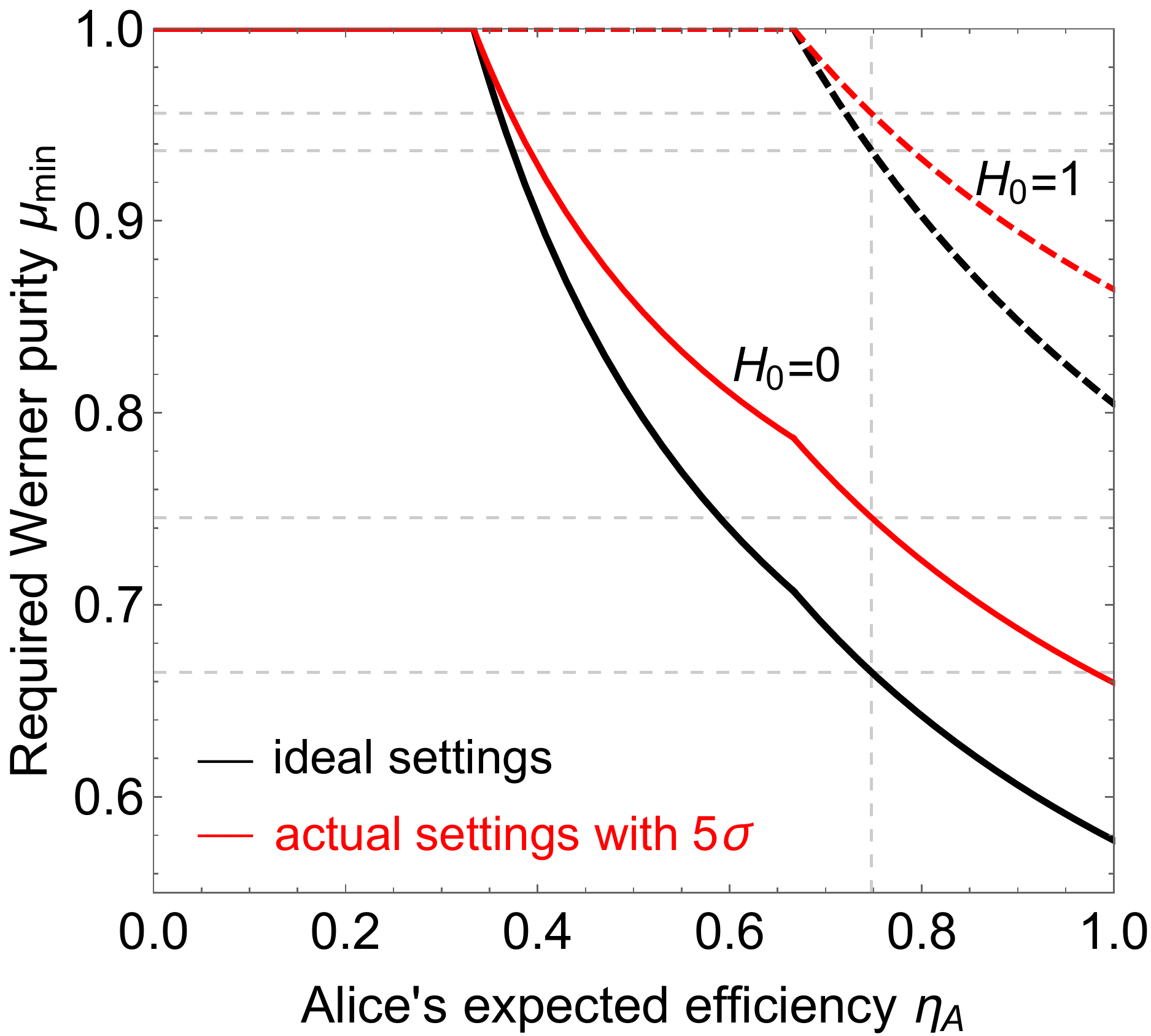}
		\caption{Loss-dependent EPR-steering bounds with  one-way communication assistance. The solid curves are the theoretical bounds for required singlet-proportion to demonstrate EPR-steering without message assistance under different measurement conditions. The dashed curves are the required purity with $H_0=1$ bit of messages assistance. The vertical dashed line represents Alice's measured efficiency $\eta_A={\left(74.8\pm 0.1\right)}\%$.}\label{werner}
	\end{center}
\end{figure}

We can now ask how good must our effective state preparation (i.e.~purity) and propagation (i.e.~loss avoidance) be if we wish to violate a communication-assisted-EPR-steering inequality. In our experiment, we  created, as closely as possible, a singlet state at the source. After distribution to Alice and Bob, this is well approximated by a loss-depleted Werner state. In Fig.~\ref{werner}, we display, as a function of Alice's efficiency, the minimum required purity $\mu_{\text{min}}={h^n_{H_0}}/\eta_A+{r_{H_0}}$ for a loss-depleted Werner state to demonstrate EPR steering when Bob's measurement axes are ideal (black curves), and with the actual settings implemented in our experiment, allowing for uncertainties (red curves) without and with $H_0=1$ message assistance. The horizontal dashed lines give the predicted minimum required singlet-proportion parameters $\mu_{\text{min}}$ for Alice's average heralding efficiency {$\eta_A=(74.8\pm 0.1)\%$} for each case. We predict that we require a minimum pure-fraction  $\mu_{\text{min}}$ of {0.745} and {0.956}, to demonstrate EPR-steering with no message and $H_0=1$ bit of messages transferred, respectively. 
Achieving a state preparation purity well above {0.956} might not seem too onerous, since visibilities $>99.6\%$  in the H/V and D/A bases have been demonstrated for Bell experiments~\cite{LHFprl2}. However it must be emphasized that the figure of  {0.956} relates to the {\em effective} state purity; imperfections in the polarization rotations by Alice's {and Bob's} Pockels cell{s} are the dominant source of effective impurity {in our experiment}.

\subsection{Results}\label{sc:results}

Here we establish the correspondence between the EPR-steering inequality~\eqref{ginequality} and the event counts in the experiment. During each trial, Alice and Bob randomly choose among one of three measurement settings then write down their own results. Alice (Bob) can record either a `+1' if she (he) observes a detection event on her (his) single H (V) detector or a `-1' if she (he) observes a detection event on her (his) single V (H) detector. In addition, Alice can record a `0' to represent no detection. {In some experimental trials both of Alice's and/or Bob's detectors click{\footnote{{Of all trials in which Bob records at least one click, Alice measures two clicks 0.26\% of the time, and Bob measures two clicks 0.24\% of the time.}}}, and Alice and Bob treat these trials differently. If both of Alice's detectors click, Alice records the outcome `0' for that trial. If both of Bob's detectors click, Bob randomly chooses to record either `+1' or `-1' as the outcome for that trial\footnote{{Intuitively, this can be understood as Bob randomly choosing one of the two photons he received to be `the' photon for that trial, and this is consistent with the outcome Bob reports when two photons are received by a single detector (which may also happen), since in that case both photons have the same outcome.}}, a procedure introduced and justified in Ref.~\cite{Smith12}.} We use `$ab|kj$' to express the trial outcome, which indicates that Alice records outcome $a$ in setting configuration $k$ and Bob records outcome $b$ in setting configuration $j$, {and we use $N_{ab|kj}$ to denote the total number of times we obtain the outcome $ab|kj$.}

{Bob's task is now to use his and Alice's data to calculate $\tilde{S}_3(r_{H_0})$, which is his experimental estimate of the steering correlation function $S_3(r_{H_0})$, Eq~\eqref{eq:genCorrFunction}. First, Bob discards data from all trials in which he and Alice measure in different bases (i.e. all trials for which $k\neq j$), as well as from all trials in which he does not detect a photon. Because Bob trusts his characterization of his measurement apparatus, and therefore his characterization of his measurement efficiencies, he can normalize the data based on these efficiencies to estimate what data would have been collected if his measurement apparatus worked with unit efficiency. Bob estimates the normalized data by computing $\mathcal{N}_{ab|jj} =  N_{ab|jj}/\eta_{B(b|j)}$. Here $\eta_{B(b|j)}$ is the efficiency with which Bob's outcome $b$ detector will click when he measures in setting $j$ --- it includes the transmission efficiency of the path photons take from the edge of Bob's lab to his polarizing beam displacer, the transmission efficiency from the output of the beam displacer to the outcome-$b$ detector, and that detector's internal efficiency. We find that these efficiencies depend on both the outcome and the measurement setting.}

{To characterize these efficiencies we modify Klyshko's method \cite {Klyshko} (which applies to systems with two photons and two detectors) to apply to our two-photon, four-detector setup. We apply this method to data that is independent from, but acquired in parallel with, the data used to test the steering inequality~\eqref{ginequality}. A full description of our method is provided in Appendix~\ref{sc:genKlyshko}. After normalizing the data in this way, Bob calculates the normalized total number of trials in which both he and Alice measure in setting $j$, given by $\mathcal{N}_{jj} = \mathcal{N}_{+1+1|jj} + \mathcal{N}_{-1+1|jj} + \mathcal{N}_{0+1|jj} + \mathcal{N}_{+1-1|jj} + \mathcal{N}_{-1-1|jj} + \mathcal{N}_{0-1|jj}$. Now, Bob is ready to calculate $\tilde{E}^c_{j}$, which is his estimate of $\langle a_j\hat{B}_j\rangle$:
\begin{equation}
\tilde{E}^c_{j} = \frac{\mathcal{N}_{+1+1|jj} + \mathcal{N}_{-1-1|jj} - \mathcal{N}_{-1+1|jj} - \mathcal{N}_{+1-1|jj}}{\mathcal{N}_{jj}},
\end{equation}
and he can also calculate $\tilde{E}^a_j$, which is his estimate of $|a_j|$:
\begin{equation}
\tilde{E}^a_j = \frac{\mathcal{N}_{+1+1|jj}+\mathcal{N}_{+1-1|jj}+\mathcal{N}_{-1+1|jj}+\mathcal{N}_{-1-1|jj}}{\mathcal{N}_{j j}}.
\end{equation}
Finally, Bob's estimate $\tilde{S}_3(r_{H_0})$, and the experimental EPR-steering inequality, are given by:
\begin{equation}
\label{vio}
\tilde{S}_3(r_{H_0}) = \frac{1}{3}\sum_{j=1}^3 \left(\tilde{E}^c_{j} - r_{H_0}\tilde{E}^a_j\right) \leq h^3_{H_0}.
\end{equation}}

Our data were taken in 90 one-minute chunks, and over the total dataset there are 936,848 experimental trials in which at least one of Bob's detectors clicks. By using the parameters in Eq.~(\ref{actrh}), the experimental residuals  between the left-hand-side and right-hand-side  of the above inequality (\ref{vio}) are {$R^3_0=0.1646\pm0.0003$} and {$R^3_1=0.0071\pm0.0003$}.  The uncertainties here represent the statistical uncertainty caused by Poissonian counting statistics. These two violations are {$478$} and {$25$} standard deviations above zero, respectively. {While we violate both inequalities by a very large number of standard deviations, an analysis method that closes the memory loophole could reduce the apparent statistical significance of this result. However, i}t is important to remember that the derivation of these inequalities includes our estimates of the uncertainty on our characterization of Bob's measurement settings up to the five-sigma level. Thus, our confidence in these two inequality violations is also at the five-sigma level (and not {478} or {25} sigma). This significant violation of the $H_0=1$ inequality conclusively demonstrates EPR-steering correlations that require more than one bit of FTL communication to simulate classically.

\section{Discussion}

Our experiment implies that if the mechanism of accounting for the quantum correlations is steering (that is, if Alice's measurement induces wave-function collapse in Bob's lab) then that mechanism requires more than one bit of information to be transmitted, faster than light, from Alice's to Bob's lab. If one is willing to accept that information is in fact being transmitted faster than light, it is interesting to ask exactly how fast this information needs to travel.
We calculate the speed at which a signal would need to travel if Alice sent it from her lab the moment her setting choice $k$ was chosen and it was received by Fenella at the spacetime point $E'$ before Bob's photon enters his lab --- this is a lower bound on the speed a signal would need to travel between Alice's QRNGs and the spacetime point $E$. (See Sec.~\ref{sc:experimentalLocalityConditions} for the distinction between these two events.) The point $E'$ is $161.3\pm1.5$ m from Alice's random number generators, and Alice's first random number generator fires $230\pm2$ ns before time $t_{E'}$. Thus, this hypothetical FTL signal would have to travel at least {$(9.84 \pm 0.14)\times 10^8$ m/s}, or {$3.28 \pm 0.05$} times the speed of light\footnote{This figure is a lower bound on the speed of the signal in the rest frame of the building containing the two laboratories. In a different reference frame, a different value would be inferred and, because the speed is superluminal, it could be anything from just above the speed of light to beyond infinite (backwards-in-time information transmission). This can be contrasted with the experiment of Salart {\em et al.}~\cite{salart08}, which gave, under certain assumptions, a lower bound on the speed $v$ of superluminal information transmission in a Bell test, regardless of the preferred frame.}.

In order to fairly evaluate the strength of the nonlocality between Alice and Bob in our experiment, we compare {$R^3_0$} and {$R^3_1$} with their corresponding Tsirelson bounds {$T^3_0$} and {$T^3_1$}. Like  the maximum quantum violation of the Clauser-Horne-Shimony-Holt Bell inequality $2\sqrt2$~\cite{tsirelson}, the Tsirelson bounds represent the maximum allowable values according to quantum theory, which can be achieved with a Bell state ($\mu=1$) and ideal projective measurements with perfect detection efficiency at Alice ($\eta_A=1$). Recall that if ${r_{H_0}}=0$, the bounds for the $H_0=0$ and $H_0=1$ inequalities are ${h^3_0}=\sqrt{3}/3$ and ${h^3_1}=(1+\sqrt{2})/3$ respectively. Thus, the maximum possible violations allowed by quantum mechanics are ${T^3_0}=(3-\sqrt{3})/3$ and ${T^3_1}=(2-\sqrt{2})/3$. Comparing our results to the maximum violations, we see that ${R^3_0/T^3_{0}}\approx0.389$ and {$R^3_1/T^3_{1}\approx0.037$}, respectively. The smallness of the latter number reflects the difficulty of demonstrating the violation of communication-assisted EPR-steering, limited mainly by Alice's total photon detection efficiency being  only ${\sim}{75}\%$.  

The conceptualization of EPR-steering as a quantum information task---to verify entanglement in the partial absence of trust---was key to its modern formulation~\cite{Howard07PRL} and its generalization to multi-party networks~\cite{He13}.  Pioneering works showed theoretically~\cite{pra_Evans} and expermentally~\cite{Bennet12,Tischler18,SA2018} that, through the use of multiple measurement settings, EPR-steering could be robust to loss even in the presence of some noise. Here, by working to minimize both noise and loss, we have been able to show the robustness of EPR steering to (hypothetical) FTL communication. Our communication-assisted and loss-tolerant EPR-steering inequalities may have applications in secure entanglement distribution with untrusted parties. In particular, in the situation where the parties are not space-like separated, adversarial communication need not be FTL  and so need not be hypothetical. {To achieve the greatest security in such schemes, it will also be important to analyze the data in a way that does not assume experimental trials are independent. Otherwise, a malicious actor could exploit the memory loophole and compromise the security of the protocol.}

It is natural to ask if one can demonstrate EPR-steering correlations requiring an even larger amount of FTL communication to simulate classically. This is certainly possible in principle. In general, if one is performing measurements on an entangled state of two qubits, one needs to perform an experiment with at least $2^{H_0}+1$ choices of measurement setting in order to exclude FTL messages of up to $H_0$ bits. As the number of measurement directions increases, some measurements necessarily become closer to each other and therefore the minimum visibility required to demonstrate an inequality violation increases. In addition, for a large enough number of settings, each measurement station will require multiple Pockels cells (or some other method of quickly switching between settings). For the specific case of ruling out messages of size $H_0=2$, five measurement settings are needed. One could choose these settings to have Bloch vectors corresponding to five of the vertices of a dodecahedron (similar to how in our three setting experiment we chose settings with Bloch vectors corresponding to three of the vertices of an octahedron). With the proper driving electronics, one could in principle use a single Pockels cell to implement fast switching between these five settings. A more significant challenge for ruling out messages of size $H_0=2$ bits is that the minimum efficiency required for such an experiment is 4/5, much higher than the 2/3 threshold needed for $H_0=1$ bits, and 
at the limit of what
has currently been demonstrated in loophole-free experiments with entangled photons~\cite{PRL2018}.
Finally, there are many open questions about the role of communication in multipartite or high-dimensional EPR-steering scenarios, where many different communication patterns can be considered.

\acknowledgments{Y.X. and Q.H. acknowledge the support from the National Natural Science Foundation of China (Grants No.~11975026, No.~61675007, and No.~12004011), the National Key R$\&$D Program of China (Grants No. 2016YFA0301302, No. 2018YFB1107205, and 2019YFA0308702), Beijing Natural Science Foundation (Grant No.~Z190005), and the Key R$\&$D Program of Guangzhou Province (Grant No.~2018B030329001). {M.W.M. acknowledges Quantum Technologies Flagship project  QRANGE (Grant Agreement No.  820405), Spanish Ministry of Science projects OCARINA (Grant No. PGC2018-097056-B-I00) and ``Severo Ochoa'' Center of Excellence CEX2019-000910-S, Generalitat de Catalunya through the CERCA program, Ag\`{e}ncia de Gesti\'{o} d'Ajuts Universitaris i de Recerca Grant No. 2017-SGR-1354, Secretaria d'Universitats i Recerca del Departament d'Empresa i Coneixement de la Generalitat de Catalunya, co-funded by the European Union Regional Development Fund within the ERDF Operational Program of Catalunya (project QuantumCat, ref. 001-P-001644), Fundaci\'{o} Privada Cellex and Fundaci\'{o} Mir-Puig.
L.K.S. acknowledges support from the National Science Foundation (RAISE-TAQS 1839223). H.M.W. acknowledges the support of the Australian Research Council Centre of Excellence Program (Grant No.~CE170100012). H.M.W. and Y.X. acknowledge the traditional owners
of the land on which this work was undertaken at Griffith
University, the Yuggera people. The authors thank Scott Glancy for helpful comments on the manuscript.}

\appendix

\setcounter{equation}{0}
\setcounter{figure}{0}
\setcounter{table}{0}
\renewcommand\thefigure{\thesection\arabic{figure}}
\renewcommand\theequation{\thesection\arabic{equation}}
\renewcommand\thetable{\thesection\Roman{table}}
\renewcommand\thesubsection{\thesection.\arabic{subsection}}
\renewcommand\thesubsubsection{\thesubsection.\alph{subsubsection}}

\section{The derivations of Eqs. (\ref{ideal0}), (\ref{s31eta1}) and (\ref{ideal1}) }

To derive the expression of Eq.~(\ref{ideal0}) in the main text, we first rewrite the bound {$h^3_0$} in this scenario. When there is no message transferred, only one state $\ket{1}_{\rm F}$ can be generated by Fenella for all three settings, and the deterministic function $\ell$ in Eq.~(\ref{eq:maxs}) is also fixed from $\{1,2,3\}$ to $\{1\}$, thus 
\beq
{h^3_0} =  \max_{\alpha} \frac{1}{3}\sum_{j=1}^{3} \delta_{a,\alpha(j)} \left(-{r_0} |a| + a \bra{1}_{\rm F} \hat{B}_{j} \ket{1}_{\rm F} \right) .
\eeq
where $\alpha$ still indicates a deterministic function from $\{1,2,3\}$ to $\{+1,0,-1\}$. It has been proven that the optimal deterministic bounds related to the second term in above equation is~\cite{pra_Evans}
\begin{equation}  \label{Gnd}
{g^3_0}=\max_{\alpha} \left[ 
\lambda_{\max} \left(\frac{1}{3} \sum_{j=1}^{3} \alpha(j) \hat{B}_{j}\right)
\right],
\end{equation}
where $\lambda_{\max}$ denotes the maximum eigenvalue of the operator within the following brackets. Then, for a lossy maximally entangled singlet state, the optimal {$r_0$} in the first term should be chosen by 
\begin{equation}\label{optr}
\underset{{r_0}}{\arg \min}  \left[  \max_{\alpha} \left( {g^{3}_{0}}- {r_0}\frac{m}{3} \right)-\eta_A(\mu-{r_0})\right].
\end{equation}
where $m$ is the total number of events when $\alpha(j)\neq0$. For example, if Alice reports null for two of the three settings, \eg, $\alpha(1)=\alpha(2)=0$, it means that $m=1$. Thus, for a given $\eta_A$, the parameter {$r_0$} and the bound {$h^3_0$} can be optimized by searching all possible cases numerically. It is reasonable that several forms of $\alpha(j)$ can give the optimized bound. For example, when $2/3\leq\eta_A<1$, both $\alpha(1)=0, \alpha(2)=\alpha(3)=+1$ and $\alpha(1)=-1, \alpha(2)=+1, \alpha(3)=0$ can get the expressions in Eq.~(\ref{ideal0}).

When we introduce the classical messages into the above scenario, the derivation will become more complicated. When $H_0=1$, there are two states $\ket{1}_\mathrm{F}$ and $\ket{2}_\mathrm{F}$ for three settings to choose. Now the bound is written as 
\beq
{h^3_1} =  \max_{\alpha,\,\ell,\,\{\ket{l}_{\rm F}\}_{l=1}^2} \frac{1}{3}\sum_{j=1}^{3} \delta_{a,\alpha(j)} \left(-{r_1} |a| + \sum_{l=1}^{2} \delta_{l,\ell(j)} a \bra{l}_{\rm F} \hat{B}_{j}\ket{l}_{\rm F} \right) ,
\eeq
where both $\ell$ and $\alpha$ are no longer fixed. Then the optimal deterministic bounds related to the second term in the above equation is
\begin{equation} 
{g^3_1}=\max_{\alpha,\,\ell} \left[\frac{1}{3} \sum_{l=1}^{2}\lambda_{\max} \left( \sum_{j=1}^{3}\delta_{l,\ell(j)} \alpha(j) \hat{B}_{j}\right)
\right],
\end{equation}
The value of {$g^3_1$} depends on both Alice's result $\alpha(j)$ and the strategy of Fenella $\ell(j)$.  

First, let us consider an idealized scenario where Alice always declares non-null results, i.e., $\alpha(j)=\pm1$. Similarly, we also need to search all possible forms for $\ell(j)$ to find the optimized $r^1$ by minimizing Eq.~(\ref{optr}). Suppose that $\ell(1)=1, \ell(2)=\ell(3)=2$, i.e., for $j=1,2,3$ measurements, the states prepared by Fenella are $\ket{1}_{\rm F}$, $\ket{2}_{\rm F}$, $\ket{2}_{\rm F}$, respectively, then we have
\begin{equation} 
{g^3_1}=\max_{\alpha} \left\lbrace \frac{1}{3}\left[\lambda_{\max} \big( \alpha(1) \hat{B}_{1}\big) +\lambda_{\max}  \big( \alpha(2) \hat{B}_{2}+\alpha(3) \hat{B}_{3}\big)\right]  \right\rbrace.
\end{equation}
By substituting $\alpha(j)=+1$ for all $j$ into this case, the optimal ${r_1}=0$ and the maximum value of {$h^3_1$} reaches $(1+\sqrt2)/3$. There are likely other combinations with different choices of $\ell(j)$ and $\alpha(j)$ that can reach the same bound in Eq.~(\ref{s31eta1}). 

Then, taking loss into account, $m$ in Eq.~(\ref{optr}) will change related to the function of $\alpha(j)$. We can still search all possible cases numerically to find the bound. Here we give one solution for the $2/3\leq\eta_A<1$ case. By keeping $\ell(1)=1, \ell(2)=\ell(3)=2$ and substituting $\alpha(1)=0, \alpha(2)=\alpha(3)=+1$, we can get the optimal expressions in Eq.~(\ref{ideal1}).

\section{Operation of Pockels cell}\label{ap:pockelscell}

We can obtain the largest violation of the steering inequality if Alice and Bob each choose three measurement settings that correspond to projective measurements in three mutually unbiased bases. Closing the locality loophole requires switching between these settings quickly. This appendix describes how we use a combination of waveplates, a single Pockels cell (PC), and a polarizing beam displacer (BD) to accomplish this.

First we define the target measurement operators we aim to measure in the experiment (at both Alice and Bob) as $\hat\sigma_z$, $\hat\sigma_x$, and $\hat\sigma_y$ for settings 1, 2, and 3, respectively. Here, $\hat\sigma_z$, $\hat\sigma_x$, and $\hat\sigma_y$ are the Pauli matrices, which are defined in the $z$-basis, $\{|0\rangle,|1\rangle\}$, as:
\begin{align}
\hat\sigma_z &= |0\rangle\langle0| - |1\rangle\langle1|,\\
\hat\sigma_x &= |+\rangle\langle+| - |-\rangle\langle-|,\\
\hat\sigma_y &= |{+i}\rangle\langle{+i}| - |{-}i\rangle\langle{-}i|,
\end{align}
where we use the convention $|\pm\rangle = \frac{1}{\sqrt{2}}(|0\rangle \pm |1\rangle)$ and $|{\pm}i\rangle = \frac{1}{\sqrt{2}}(|0\rangle \pm i |1\rangle)$. Using the Bloch representation, we display the `+' eigenstates of the three measurement settings in Fig.~\ref{fg:targetMeas}.

The states in our experiment are encoded in polarization, and thus we must map the Pauli operators to measurements on the polarization degree of freedom. We map the $z$-basis to horizontal ($|H\rangle$) and vertical ($|V\rangle$) polarization with the assignments $|0\rangle \rightarrow |V\rangle$ and $|1\rangle \rightarrow |H\rangle$. As a result, the $x$-basis is mapped to the diagonal and anti-diagonal polarization basis via $|{+}\rangle\rightarrow |D\rangle$ and $|{-}\rangle\rightarrow |A\rangle$, and the $y$-basis is mapped to the left- and right-circular polarizations via $|{+i}\rangle\rightarrow |L\rangle$ and $|{-i}\rangle\rightarrow |R\rangle$. 
\begin{figure}
\includegraphics[width=\columnwidth]{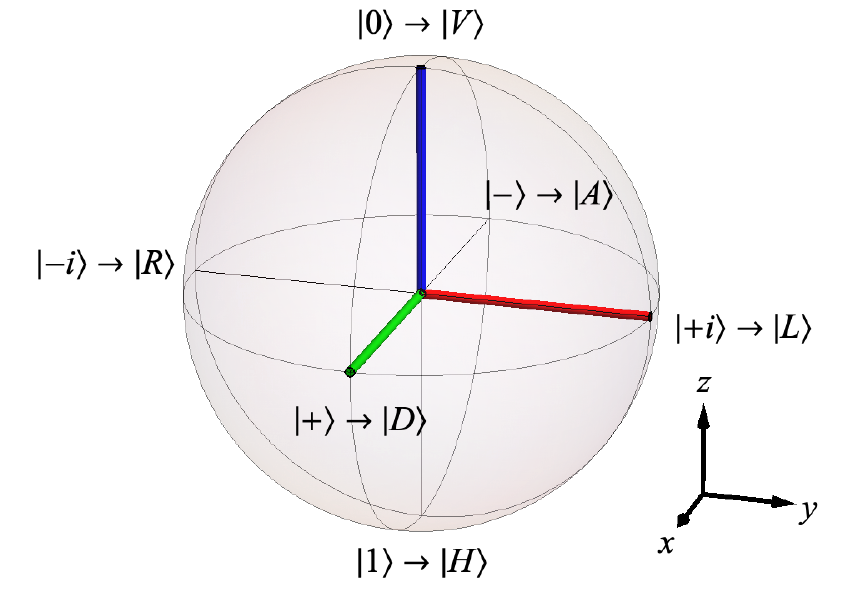}
\caption{Bloch sphere representation of target measurements in our experiment. The blue, green, and red Bloch vectors represent the `+' eigenstates of the target measurements for settings 1, 2, and 3 respectively. The figure also denotes how we map the Bloch sphere to polarization states. As a result of our choice of coordinate system, vertically polarized light $|V\rangle$ is represented by the Bloch vector $(0,0,1)$, diagonally polarized light has Bloch vector $(1,0,0)$, and left-circularly polarized light has Bloch vector $(0,1,0)$.}
\label{fg:targetMeas}
\end{figure}

To realize the target measurements, we need a scheme for which the overall effect of Bob's measurement is that, for setting $j=1$, polarization state $|V\rangle$ remains $|V\rangle$, for $j=2$, $|D\rangle$ is rotated to $|V\rangle$, and for $j=3$, $|L\rangle$ becomes $|V\rangle$. It is easiest to describe our scheme if we consider how a vertically polarized beam is affected as it travels \emph{backwards} through the optics in Bob's measurement setup, beginning from the `+' detector. For the $j$-th measurement setting, vertically polarized light will be transformed to some other polarization state as it propagates backwards, and the polarization state that the beam ends up in is the `+' eigenstate of this $j$-th measurement.

We represent the polarization of the light beam at various stages of backwards-propagation in Fig.~\ref{fg:pcScheme}.  Initially, the light transmitted by the beam displacer is vertically polarized (Fig.~\ref{fg:pcScheme}(a)). Next the beam propagates backwards through two waveplates, and it is rotated to an elliptical polarization state (Fig.~\ref{fg:pcScheme}(c)). Then the beam encounters the PC. Ideally, when no voltage is applied to the PC, it has no birefringence and thus has no effect on polarization. The PC becomes birefringent when a voltage is applied, and it is aligned such that it induces a relative phase $\theta$ between the $|D\rangle$ and $|A\rangle$ polarization states. In our setup, setting $j=1$ corresponds to no voltage applied to the PC, $j=2$ corresponds to a positive voltage which applies a phase of $+\theta$, and $j=3$ corresponds to a negative voltage that applies a phase $-\theta$. We choose $\theta=120^\circ$, which ensures the three different polarization states exiting the PC are eigenstates of three different mutually unbiased measurement operators. Finally, the light backwards-propagates through the final QWP and HWP and we can see that the three resulting Bloch vectors (Fig.~\ref{fg:pcScheme}(f)) are equal to the Bloch vectors of the `+' eigenstates of the three target measurement operators (Fig.~\ref{fg:pcScheme}).

\begin{figure*}
\includegraphics[width=\textwidth]{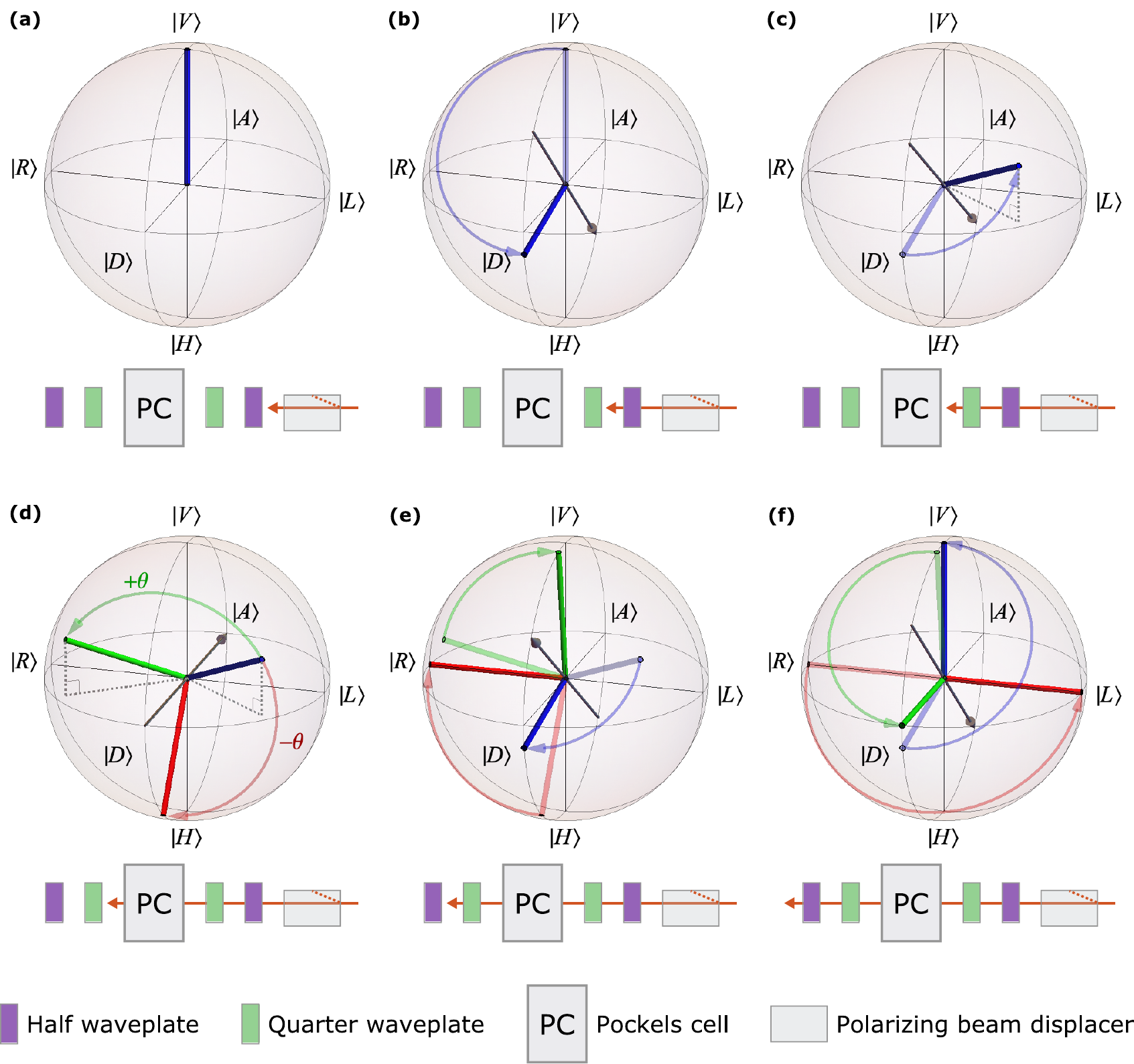}
\caption{Visualization of polarization rotation light experiences as it propagates backwards through Bob's measurement setup. In each subfigure we plot the beam's polarization state on the Bloch sphere, after it has propagated backwards through part of Bob's setup. (a) After propagating through the beam displacer, the beam is vertically polarized, as represented by the blue Bloch vector. Next the polarization state is rotated by (b) a HWP with fast-axis $-24.94^\circ$ from horizontal and (c) a QWP at an angle of $-27.38^\circ$. The grey vector in the linear-polarization plane of the Bloch sphere indicates the angle of the waveplate; the action of a half (quarter) waveplate is to rotate the Bloch vector by an angle of $180^\circ$ ($90^\circ$) about this grey vector. Since we are considering \emph{backwards} propagation, the angle of rotation is determined by the \emph{left}-hand rule (the Bloch vector rotates in the direction of the curl of the fingers of the left hand when the left thumb points in a direction parallel to the grey vector). (d) When the beam travels through the PC, one of three rotations are applied to the beam, depending on the measurement setting. We either apply no rotation to the Bloch vector ($j=0$, blue Bloch vector), apply a rotation of $+\theta$ ($j=1$, green Bloch vector), or apply a $-\theta$ rotation ($j=2$, red vector). Each of the three resulting Bloch vectors is orthogonal to the other two, indicating that they each correspond to the `+' eigenstate of one of a set of three mutually unbiased bases. Finally the beam travels through (e) a QWP at $62.62^\circ$ and (f) a HWP at $-24.94^\circ$. The overall effect of backwards propagation through Bob's measurement is that, if $j=0$, the polarization state $|V\rangle$ remains $|V\rangle$, if $j=1$, $|V\rangle$ is rotated to $|D\rangle$, and when $j=2$, $|V\rangle$ becomes $|L\rangle$.}
\label{fg:pcScheme}
\end{figure*}

The above scheme works for an idealized PC. In reality, the PC's we used in Alice and Bob's measurement stations exhibited a small amount of birefringence even when no voltage was applied to them. To correct for this, we performed quantum process tomography to characterize the Jones matrix of each PC when no voltage was applied. Instead of using the idealized waveplate angles given in the caption of Fig.~\ref{fg:pcScheme}, we used a numerical optimization to find the optimal angles for each measurement station.

\section{Quantum characterization of Bob's measurements}\label{sc:BobMeas}
\setcounter{figure}{0}

In order to perform the steering test, we need an accurate quantum description of the measurements performed in Bob's laboratory for each of the three settings, and we detail how we find this description in this appendix. We begin by defining the family of quantum models we use to represent Bob's measurement apparatus in subsection~\ref{sc:measQM} of this appendix. We perform two types of measurements in order to learn a model that accurately represents Bob's apparatus. First, we characterize the polarization-rotation optics in Bob's measurement setup by performing a bootstrapped version of standard quantum measurement tomography with parametric resampling. We explain our tomography procedure in subsection~\ref{sc:PBQMT}, and we explain how we modify the steering inequalities to deal with imperfections in Bob's measurement settings in subsection~\ref{sc:actualMeasTheory}. Second, we measure the path efficiencies for each of Bob's detection outcomes (and track these efficiencies in real time) using a modified version of Klyshko's efficiency measurement, which we detail in subsection~\ref{sc:genKlyshko}. 

\subsection{Quantum model of Bob's measurement apparatus}\label{sc:measQM}

A simplified diagram of our experiment is given in Fig.~\ref{fg:eff_setup}(a). Bob's measurement setup consists of two waveplates, followed by a PC, followed by two more waveplates, and a polarizing beam displacer (BD) with a detector in each output port. We model the combined effect of the waveplates and PC as a polarization rotation that depends on Bob's choice of measurement setting, $j$. The BD splits the beam into two paths; vertically polarized light is directly transmitted into the `+' path, and horizontally polarized light is deflected into the `-' path. Bob records a `+' outcome if the detector in the `+' path clicks, a `-' outcome if the other detector clicks, and a null outcome if neither detector clicks. To account for non-unit detection efficiency we model the detectors as ideal detectors preceded by non-polarizing beamsplitters; the transmission probability of the beamsplitter in the `$\pm$' path is given by $\beta_{j}^{(\pm)}$.

\begin{figure}
\includegraphics[width=\columnwidth]{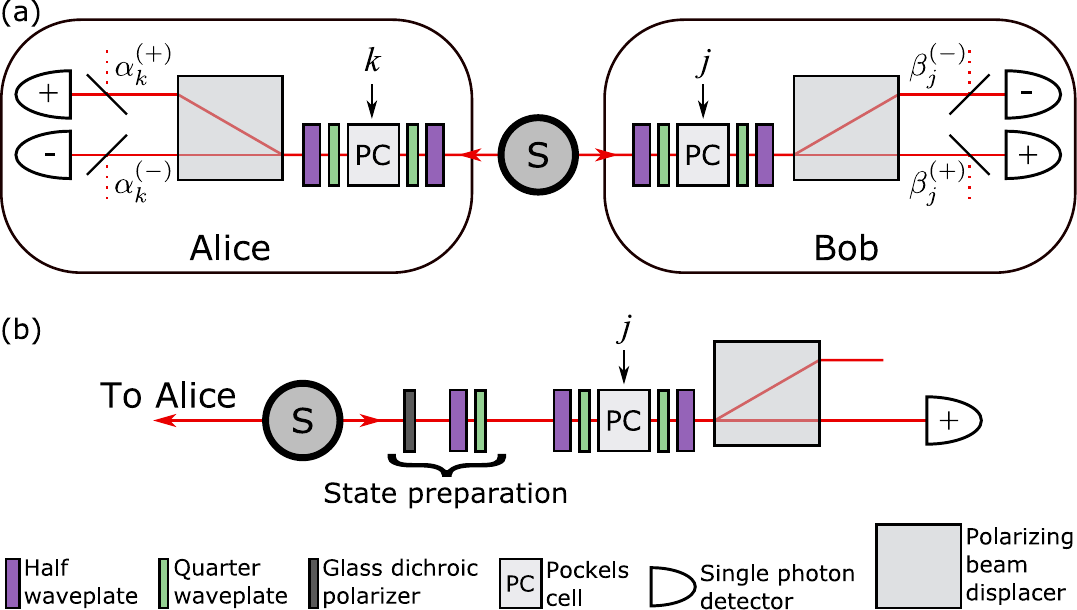}
\caption{Simplified diagrams of the experimental setups used for (a) testing the steering inequality and (b) performing tomography on Bob's measurement apparatus. (a) A source emits a pair of photons; one travels to Alice and one to Bob. At each measurement station, photons propagate through polarization-rotating optics, including a PC with an applied voltage that depends on the measurement settings. In this simplified model, we assume there is no polarization-dependent loss before the light encounters a polarizing beam displacer (BD). We model loss in the experiment by first modeling all optics and detectors with perfect efficiency, and introducing non-polarizing beamsplitters with setting-dependent transmission probabilities $\alpha_k^{(+)}$, $\alpha_k^{(-)}$, $\beta_j^{(+)}$, and $\beta_j^{(-)}$. (b) We add a polarizer and two waveplates at Bob's station to perform the measurements required for quantum measurement tomography.}
\label{fg:eff_setup}
\end{figure}

In our model of Bob's measurements, we will make two main assumptions. First, we assume that in Bob's measurement apparatus there is no polarization-dependent loss before the BD; this allows us to model all the loss in the experiment as occurring after the BD. When characterizing Bob's measurement apparatus, we find no significant evidence of polarization dependent loss in the waveplates and Pockels cell. Second, we assume that Bob's measurements are lossy versions of projective measurements. We measure an excellent extinction ratio greater than $10^5{:}1$ with the beam displacer, providing evidence in favour of this assumption. We note that this second assumption is conservative, because if it were the case that Alice and Bob \emph{do not} share an entangled state, then the bound of the steering inequality can only be achieved if $E_{j}^{(\pm)}$ are projectors.

In quantum theory, the positive-operator valued measure (POVM) representating Bob's three-outcome measurement for setting $j$, is:
\begin{equation}\label{eq:bobPOVM}
M_j^B = \left\{\beta_{j}^{(+)}\hat E_{j}^{(+)},\,\beta_{j}^{(-)}\hat E_{j}^{(-)},\,\hat E_{j}^{(\mathrm{null})} \right\}.
\end{equation}
The measurement effects satisfy $\beta_{j}^{(+)} \hat E_{j}^{(+)} + \beta_{j}^{(-)}\hat E_{j}^{(-)} + \hat E_{j}^{(\mathrm{null})} =\mathbb{I}$, where $\mathbb{I}$ is the $2\times2$ identity matrix. For a measurement of a single photon with polarization state $\rho$, the probability that detector $\pm$ clicks is $P(\pm|\rho,j) = \beta_{j}^{(\pm)}\Tr\left(\rho \hat E_{j}^{(\pm)}\right)$, and the probability that neither detector clicks is $P(\mathrm{null}|\rho,j) = \Tr\left(\rho \hat E_{j}^{(\mathrm{null})}\right)$. We represent the measurement effect operators $\hat E_{j}^{(\pm)}$ as:
\begin{equation}
E_{j}^{(\pm)} = \frac{1}{2}\left(e_{j,0}^{(\pm)}\mathbb{I} + \bm{e}_{j}^{(\pm)} \cdot \bm\sigma \right),
\end{equation}
where $\bm\sigma = (\hat\sigma_x,\hat\sigma_y,\hat\sigma_z)$ is a vector of Pauli matrices, and $\bm e_{j}^{(\pm)} = \left(e_{j,1}^{(\pm)},e_{j,2}^{(\pm)},e_{j,3}^{(\pm)}\right)$ is the Bloch vector consisting of three real numbers that specify the basis of the measurement. Because of our assumption that Bob's measurements are lossy projective measurements, we set $e_{j,0}=1$, and the Bloch vectors $\bm e_{j}^{(\pm)}$ must satisfy $\left|\bm e_{j}^{(\pm)} \right| = 1$. Finally, since in our model we assume there is no polarization-dependent loss before the BD,
the measurement effects must satisfy $\hat E_{j}^{(+)} + \hat E_{j}^{(-)} = \mathbb{I}$, implying $\bm e_{j}^{(+)} = -\bm e_{j}^{(-)}$. Thus, in order to fully specify measurement $M_j^B$ it is sufficient to know the Bloch vector $\bm e_{j}^{(+)}$, and the two efficiencies $\beta_{j}^{(+)}$ and $\beta_{j}^{(-)}$. 

\subsection{Bootstrapped measurement tomography with parametric resampling}\label{sc:PBQMT}

We slightly modify our setup to perform tomography on Bob's measurement device (see Fig.~\ref{fg:eff_setup}(b)). We tune the source to produce the separable state $\left|VH\right\rangle$, so that a vertically polarized photon is sent to Alice's measurement station, and a horizontally polarized one to Bob's. We place a horizontal polarizer in Bob's measurement bridge, followed by a half and a quarter waveplate which we can rotate to prepare one of the six polarization states $\left|H\right\rangle$, $\left|V\right\rangle$,  $\left|D\right\rangle$, $\left|A\right\rangle$, $\left|R\right\rangle$, or $\left|L\right\rangle$. We turn Alice's PC off and rotate her waveplates to maximize the rate of detections in her `+' detector. We set Bob's PC to randomly switch between his three measurement settings, and acquire data for one minute. The PC runs at a rate of 100 kHz, such that Bob performs roughly $3.3\times10^{4}$ measurement trials per second, per measurement setting. We acquire data over a total of 120 one-minute intervals, and each minute we set the input state to one of the above six polarization states. For each minute of data we record the input polarization state, the number of coincident detections recorded between Alice's and Bob's `+' detectors while Bob's PC was set to each of the three measurement settings, as well as the number of trials the PC was set to each setting. For each setting, the final dataset is a list of 120 measured pairs of input states and coincidence count rates.

We analyze the data for each setting independently, using a bootstrap method with parametric resampling. For each measurement setting $j$ we perform 10,000 analysis trials. Each analysis trial begins by sampling, with replacement, 120 pairs of input states and coincidence rates; this is the bootstrapping. Next, to account for statistical errors in the count rates we measure, as well as systematic errors caused by imperfections in the waveplates used to prepare the input states, we perform a parametric resampling step. We resample each measured count rate from a Poisson distribution with mean equal to the raw count rate that we measured. Based on the manufacturer's specified tolerance, we assume that the phase of each state-preparation waveplate is accurate to within $\pm 0.005$ wavelengths, and we choose an error value for each waveplate from a uniform distribution within this range. We also assume that we can find the zero angle for each waveplate to within an accuracy of 0.1 degrees, and we choose an error value from a normal distribution with standard deviation 0.1 for each waveplate. Then we recalculate the six input states based on the randomly assigned waveplate imperfections.
After both the bootstrapping and parametric resampling steps we are left with a new list of data consisting of 120 (imperfect) input states and resampled coincidence count rates. Finally, using this new list of data we find $\tilde{\bm e}_{j}^{(+)}$, the maximum-likelihood estimate of $\bm e_{j}^{(+)}$. Because we are making the conservative assumption that the $E_{j}^{(\pm)}$ are projectors, we enforce the constraint $\left|\tilde{\bm e}_j^{(+)}\right| = 1$.

After performing all analysis trials, we are left with a list of 10,000 normalized maximum likelihood estimates of $\bm e_{j}^{(+)}$. The mean of this list, $\langle \tilde{\bm e}_{j}^{(+)} \rangle$, is our final estimate of $\bm e_{j}^{(+)}$. When the 10,000 max-likelihood estimates are plotted, their endpoints lie on the surface of the Bloch sphere, and they are distributed around $\langle \tilde{\bm e}_{j}^{(+)} \rangle$ in the shape of a 2-D Gaussian ellipse.  We calculate $\sigma_j$, the standard deviation of angular displacements from the mean vector in the direction along the semi-major axis of this ellipse. The results of the measurement tomography analysis are summarized in Table~\ref{table1} in the main text. When testing the steering inequality, we assume that the angle between the true Bloch vector describing Bob's measurement, $\bm e_{j}^{(+)}$, and our best-guess vector $\langle \tilde{\bm e}_{j}^{(+)} \rangle$ is no larger than $5\sigma_j$. 

\subsection{Dealing with imperfections in Bob's measurements}\label{sc:actualMeasTheory}

In order for Bob to make the most conservative characterization of his measurements, we make Bob's measurement Bloch vectors as close as possible after the rotations by corresponding angles with $5\sigma$ uncertainties. The idea is that if Bob's settings were all closer together (i.e. closer to a ``common'' direction than he thought they were), then Alice could convince him that she was steering him even if she wasn't. At the extreme, if Bob's Pockels cell stopped working so that he was always measuring the same direction, then Alice could, in each run, simply dial up which result ($+1$ or $-1$) she wanted Bob to get by having Fenella send a photon aligned or anti-aligned with this direction. She could get any correlation she desired with no entanglement at all. This is why we consider the worst case for Bob's directions, when they are all closer to a common direction.

For the no message case, Alice/Fenella's best strategy is to prepare a state aligned/anti-aligned with the mean (normalized) direction $\vec{m}$ of Bob's three measurement settings, then send it to Bob. We can flip the measurement directions $\vec{B}_j$ (\eg, from $\vec{B}_1$ to its opposite $-\vec{B}_1$) and the results (from $+1$ to $-1$) without changing anything physically. Depending on the sign of each $\vec{B}_j$ we choose, there are $8$ different combinations and their respective normalized mean vectors. We need to find, out of these $8$ groups, the one that will have the largest inner product between their own means and each of the $\vec{B}_j$ that go into it--that is $\{-\vec{B}_1, \vec{B}_2, \vec{B}_3\}$. Having identified this, we can rotate each $\{-\vec{B}_1, \vec{B}_2, \vec{B}_3\}$ towards $\vec{m}=(-\vec{B}_1+\vec{B}_2+\vec{B}_3)/|-\vec{B}_1+\vec{B}_2+\vec{B}_3|$ by their respective uncertainties ($5\sigma_j$) through the Rodrigues formula. 

\begin{table}[!htb]
\centering 
	\renewcommand\arraystretch{1.25}
	\caption{Conservative estimate of Bob's measurements after rotations by $5\sigma$ uncertainties.}\label{brot}
	\begin{tabular*}{\columnwidth}{ c@{\extracolsep{\fill}} c S[table-format=2.4] S[table-format=2.4] S[table-format=2.4] }
\hline\hline\Tstrut
{}\Tstrut & {} & {Setting 1}& {Setting 2}& {Setting 3}\\
\hline
		\multirow{3}*{$H_0=0$} &X\Tstrut &0.0913 & 0.9942 & 0.1424  \\
		&Y\Tstrut & -0.0024 & 0.0943 & 0.9875  \\
		&Z\Tstrut & -0.9958 & -0.0508 & -0.0671 \\
\hline
		\multirow{3}*{$H_0=1$} &X\Tstrut & {} & 0.9936 & 0.1584 \\
		&Y\Tstrut & {} & 0.1127 & 0.9869 \\
		&Z\Tstrut & {} & -0.0104 & -0.0278 \\
\hline\hline
\end{tabular*}
\end{table}

When an $H_0=1$ bit message is transferred, Alice has $d=2$ level of
information that can encode something about the measurement
setting Bob will choose. For instance, we assume that Alice's
first level of information indicates that Bob will measure along $\vec{B}_3$, and the second level message means that Bob will measure either $\vec{B}_1$ or $\vec{B}_2$ (each with 50\% probability).
In this situation, Alice's optimal cheating strategy would be to ask Fenella to prepare the state $\ket{1}_{\rm F}$ aligned with $\vec{B}_3$ if her message is on the first level, and to prepare the state $\ket{2}_{\rm{F}}$ along $\vec{m}=(\vec{B}_1+\vec{B}_2)/|\vec{B}_1+\vec{B}_2|$ if her message is on the second level.
Now, to avoid Alice's optimal cheating strategy, Bob should find the closest two of the three settings among 12 different combinations of $\{\pm\vec{B}_u, \pm\vec{B}_v\}$ ($u,~v\in \{1,2,3\}$). After checking their respective inner products, we find $\vec{B}_2$ and $\vec{B}_3$ are closest to each other, then rotate them towards their mean vector $(\vec{B}_2+\vec{B}_3)/|\vec{B}_2+\vec{B}_3|$ (each by $5\sigma_j$). Note that in this case it doesn't matter how much Bob rotates his $\vec{B}_1$ by, because Alice/Fenella can always prepare a state that perfectly aligns with $\vec{B}_1$. The results of Bob's measurement settings after rotation are summarized in Table \ref{brot}. The optimal parameter $r$ and bounds $h_n^{H_0}$ are obtained based on these rotated settings.

\subsection{Measurement of Bob's detection efficiencies}\label{sc:genKlyshko}

David Klyshko invented a widely used method to estimate overall detection efficiency in experiments with two correlated photons that each travel to one detector~\cite{Klyshko}. In our experiment each photon can travel to one of {\em two} detectors, and we modify Klyshko's method to allow us to estimate the overall detection for efficiency for each of the four detectors in this type of setup.

\subsubsection{Modified Klyshko equations}

To explain our method, we refer again to the simplified model of our experiment in Fig.~\ref{fg:eff_setup}(a). Our goal is to characterize the efficiencies $\beta_{j}^{(+)}$ and $\beta_{j}^{(-)}$. We do this by first deriving equations for the count rates Alice and Bob expect to observe. For now, we will assume that any loss in the setup affects both polarizations of light equally. We will revisit this assumption later. Each photon pair emitted by the source may take one of four paths towards the detectors at Alice's and Bob's measurement stations; Alice's and Bob's photons might each travel towards a `+' detector, Alice's can travel to her `+' detector and Bob's to his `-' detector, Alice's could travel to her `-' detector and Bob's to his `+' detector, or finally both photons can travel towards the `-' detectors. Each of these four paths has an associated joint probability represented by $p_{k.j}^{a,b}$, with $a,b \in \{+,-\}$, representing Alice's and Bob's outcomes, and $k,j \in \{1,2,3\}$ representing their settings.\footnote{Another way to understand the meaning of the joint probabilities  $p_{k.j}^{a,b}$ is by considering the case of an experiment with no loss (i.e. consider $\alpha_{k}^{(\pm)} = \beta_{j}^{(\pm)} = 1$); in this case, $p^{(a,b)}_{k,j}$ is the probability that, in an experiment in which Alice and Bob measure a single pair of photons with settings $k$ and $j$, respectively, Alice obtains outcome $a$ and Bob obtains outcome $b$. In a real experiment with non-zero loss, the probability of Alice and Bob obtaining the joint outcome $(a,b)$ is actually $\alpha_{k}^{(a)}\beta_{j}^{(b)}p_{k,j}^{(a,b)}$. In principle the probabilities $p_{k,j}^{(a,b)}$ could be computed given a complete quantum description of Alice and Bob's measurement apparatuses as well as the state shared between them. However while the $p_{k,j}^{(a,b)}$'s are a useful concept for deriving the modified Klyshko equations, they do not need to be known in order to find the efficiencies $\alpha_{k}^{(a)}$ and $\beta_{j}^{(b)}$.}

Using these joint probabilities, we can write down the expected detection rates if Alice and Bob measure light from a source that produces $N$ pairs each second. We assume that the source emits at most one pair per coincidence window, and we also assume that the detectors have negligibly small rates of background counts. We will also revisit these two assumptions later. With this, the expected rates of coincident detections $C_{k,j}^{(a,b)}$ between detector $a$ at Alice and detector $b$ at Bob are:
\begin{align}
C_{k,j}^{(a,b)} &= N \alpha_{k}^{(a)} \beta_{j}^{(b)}p_{k,j}^{(a,b)}. \label{eq:Cab}
\end{align}
The expected rates of single detections at detector $a$ at Alice, $A_{k}^{(a)}$, are:
\begin{align}
A_{k}^{(a)} &= N \alpha_{k}^{(a)}\left(p_{k,j}^{(a,+)}+p_{k,j}^{(a,-)}\right). \label{eq:Aa}
\end{align}
Finally, the expected rates of single detections at detector $b$ at Bob, $B_j^{(b)}$, are:
\begin{align}
B_{j}^{(b)} &= N \beta_{j}^{(b)}\left(p_{k,j}^{(+,b)}+p_{k,j}^{(-,b)}\right).\label{eq:Bb}
\end{align}

Rearranging Eqs.~\eqref{eq:Cab}-\eqref{eq:Bb} we find that Alice's efficiency $\alpha_{k}^{(a)}$ is given by:
\begin{align}
\alpha_{k}^{(a)} &= \frac{C_{k,j}^{(a,\bar a)}C_{k,j}^{(\bar a,a)} - C_{k,j}^{(a, a)}C_{k,j}^{(\bar a, \bar a)} }{C_{k,j}^{(\bar a, a)}B_j^{(\bar a)}-C_{k,j}^{(\bar a, \bar a)}B_j^{(a)}}, \label{eq:etaAa} 
\end{align}
and Bob's efficiency $\beta_j^{(b)}$ by:
\begin{align}
\beta_{j}^{(b)} &= \frac{C_{k,j}^{(b,\bar b)}C_{k,j}^{(\bar b,b)} - C_{k,j}^{(b, b)}C_{k,j}^{(\bar b, \bar b)} }{C_{k,j}^{(b,\bar b)}A_k^{(\bar b)}-C_{k,j}^{(\bar b, \bar b)}A_k^{(b)}}, \label{eq:etaBb}
\end{align}
where the overbars {on the $a$'s and $b$'s} in the superscripts {in the above two equations} denote ``the opposite outcome'' (e.g. if $b=\,$`$-$', then $\bar b =\,$`+'). Equations~\ref{eq:etaAa} and~\ref{eq:etaBb} provide a method for computing the efficiency of each of Alice's and Bob's detection outcomes from the pair source to each detector, using only the measured singles and coincidence rates in the experiment. Importantly, the probabilities $p_{k,j}^{(a,b)}$ all factor out, and thus we can characterize the efficiencies of all four detectors without having details of the setup such as the quantum state of the photon pairs or the measurement operators describing Alice's and Bob's measurement stations.

While the accuracy of Eqs.~\eqref{eq:etaAa}-\eqref{eq:etaBb}, does not depend on the state of the photon pair or on the specific measurements performed, the precision does---this is due to the subtraction of two terms in the numerators and denominators. Specifically, these equations are most precise when Alice's and Bob's measurement outcomes are strongly correlated (or anticorrelated) and least precise when their outcomes are uncorrelated. For example, if the source emits the maximally entangled singlet state and Alice and Bob perform measurements in the same basis, then the positive term in each of the numerators and denominators will be much larger than the negative one, and Eqs.~\eqref{eq:etaAa}-\eqref{eq:etaBb} will be a ratio of two large numbers with error bars determined by Poissonian counting statistics. On the other hand, if Alice and Bob measure the singlet state in two mutually unbiased bases, then both terms in the numerators and denominators will be of similar orders of magnitude, and the numerators and denominators will both approach 0, leading to a larger overall uncertainty in the final estimate of the efficiencies.

In order to test Eq.~\eqref{vio} it suffices to know the {\em ratio} of efficiencies of Bob's two detectors. To get this ratio, we divide the $b=\,$`$+$' form of Eq.~\eqref{eq:etaBb} by its $b=\,$`$-$' form. The detector outcomes between Alice and Bob are maximally anticorrelated---and the modified Klysho equations are most precise---when Alice measures in the same setting as Bob ($k=j$). Thus, to find the ratio of the detection efficiencies for each of Bob's outcomes, we use the formula:
\begin{equation}
\beta_{j}^\mathrm{ratio} = \frac{\beta_{j}^{(+)} }{\beta_{j}^{(-)} } = \frac{C_{j,j}^{(-,+)}A_{j}^{(+)}-C_{j,j}^{(+,+)}A_{j}^{(-)}}{C_{j,j}^{(+,-)}A_{j}^{(-)}-C_{j,j}^{(-,-)}A_{j}^{(+)}}. \label{eq:Bobeffratio}
\end{equation}

We will now revisit the assumption that any lossy components of our experimental setup affect both polarizations of light equally. When characterizing our experiment, we do not find any evidence of polarization-dependent loss through any of the individual optics in our experiment. However, there is still some polarization-dependent loss at the source. Each photon produced in the source is split into two paths (the $|H\rangle$ component of the photon travels down one path, and the $|V\rangle$ component down another other), and the efficiency with which these two paths are coupled into the single-mode fibers leading to Alice's and Bob's measurements stations are not equal. To account for this, we simulate the count rates we would expect to measure in the presence of some polarization-dependent loss, and see what size of systematic error these simulated rates introduce into the equation for $\beta^\mathrm{ratio}_j$. We find that if one polarization experiences $x\%$ more loss than the orthogonal polarization, then this can introduce an error of at most about $\pm x\%$ in Eq.~\ref{eq:Bobeffratio}. The exact amount of systematic error introduced depends on the bases that Alice and Bob measure in. If we can bound the polarization-dependent loss in the setup to be below a specific level, we can also bound the systematic error in our estimate of $\beta^\mathrm{ratio}_j$ to that same level.

\subsubsection{Efficiency measurements}

During our experiment, each time a voltage is applied to the PC it is applied for approximately 150 ns, which is enough time for 12 laser pulses to pass through the PC. We record data for all of these pulses, but only use the data from one pulse to test our steering inequality. We call this pulse the {\em inequality pulse}. We average the data from the other 11 pulses (which we call {\em calibration pulses}) and use these averaged data to estimate the efficiency of Bob's measurements. Using the calibration pulses allows us to keep track of $\beta_{j}^\mathrm{ratio}$ in real time while testing the steering inequality. We use Eq.~\eqref{eq:Bobeffratio} to find $\beta_{j}^\mathrm{ratio}$ for each of Bob's three measurement settings.

\begin{figure}
\includegraphics[width=\columnwidth]{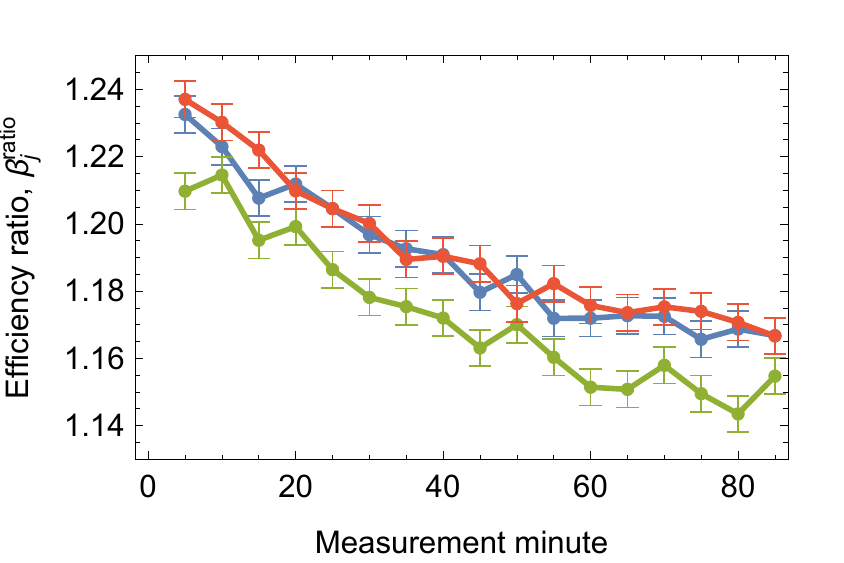}
\caption{Ratio of system detection efficiency between Bob's ${+}1$ and ${-}1$ measurement outcomes throughout the running time of the experiment, for measurement settings $j{=}1$ (blue), $j{=}2$ (green), and $j{=}3$ (orange-red). Each point is calculated from five minutes of data, and error bars represent a 1-sigma statistical uncertainty estimated from the Allan deviation of the data.}
\label{fg:bobeffratios}
\end{figure}

\begin{table}
	\renewcommand\arraystretch{1.25}
\caption{Measurement and 5-sigma conservative estimate of $\beta_{j}^\mathrm{ratio}$ for each measurement setting.}
\label{tab:bobeffratios}
\begin{tabularx}{\columnwidth}{ >{\centering\arraybackslash}X >{\centering\arraybackslash}X >{\centering\arraybackslash}X }
\hline\hline
Setting & Measured value closest to 1 & Conservative estimate \\ [0.5ex] 
 \hline
 1 & $1.1657 \pm 0.0054$ & 1.115  \\ 
 2 & $1.1435 \pm 0.0054$ & 1.094  \\
 3 & $1.1666 \pm 0.0054$ & 1.116  \\
  \hline\hline
\end{tabularx}
\end{table}

Calibration pulses were collected in 85 separate one-minute intervals\footnote{We collected 90 one-minute sets of inequality-pulse data, however due to an error in recording the data we only collected calibration data for 85 of those minutes.}.
Calculating the Allan deviation~\cite{Allan} of the data implies that the smallest uncertainty on $\beta_{j}^\mathrm{ratio}$ is obtained if the data is averaged over intervals of approximately five minutes. The total dataset can be divided into 17 five-minute windows. To visualize how $\beta_{j}^\mathrm{ratio}$ varies over the running time of the experiment, we calculate it for each of the five-minute windows, and plot this is in Fig.~\ref{fg:bobeffratios}. We notice a clear difference in $\beta_{j}^\mathrm{ratio}$ between setting 2 and settings 1 and 3. We also notice a significant drift in $\beta_{j}^\mathrm{ratio}$ throughout the experiment.
The $\beta^{\mathrm{ratio}}_j$'s we measure are all greater than one. For the specific $H_0 = 0$ and $H_0 = 1$ inequalities we test, it becomes easier to violate the inequality as $\beta^{\mathrm{ratio}}_j$ increases. Thus, the conservative choice is to underestimate the value of each $\beta^{\mathrm{ratio}}_j$. We are confident that the amount of polarization-dependent loss in the source is no larger than $2\%$. For each setting we conservatively estimate $\beta_{j}^\mathrm{ratio}$ as the lowest measured value from among the five-minute windows, minus the maximum $2\%$ systematic error we expect from polarization-dependent loss, minus five times the estimated standard error on that measurement. The conservative estimates of $\beta_{j}^\mathrm{ratio}$ we use to test the steering inequality are listed in Table~\ref{tab:bobeffratios}.

\subsubsection{Systematic uncertainties}

There are two other types of systematic error that Eq.~\eqref{eq:Bobeffratio} does not take into account, namely background counts from the detectors and multi-pair emissions from the source. We measure a background rate of approximately 200 counts/second, and the probability that any single laser pulse creates a photon pair is about 0.0072, which means that the probability of creating two pairs is $0.0072^2 \approx 5\times 10^{-5}$. We simulate more realistic detection rates that take the background counts and multi-pair emission probability into account, and after evaluating Eq.~\eqref{eq:Bobeffratio} with these simulated rates we see that these systematic errors lead to an {\em underestimation} of $\beta_{j}^\mathrm{ratio}$ by approximately 0.0006. Because these two sources of error combine to introduce a systematic uncertainty that is roughly 10\% the magnitude of the statistical uncertainty on our measurement of $\beta_{j}^\mathrm{ratio}$, and also because they lead to an even more conservative estimate of $\beta_{j}^\mathrm{ratio}$, we can safely neglect them.

%%%%%%%%%%%%%%% References %%%%%%%%%%%%%%%%%%%%%%%%%

\end{document}